\documentclass[11pt,journal,compsoc]{IEEEtran}
\usepackage{comment}

\ifCLASSOPTIONcompsoc
  \usepackage[nocompress]{cite}
\else
  \usepackage{cite}
\fi
\usepackage{algorithm}
\usepackage{algorithmic}

\ifCLASSINFOpdf
  \usepackage[pdftex]{graphicx}
\else
\fi
\usepackage[hyphens]{url}

\usepackage{amsmath}
\usepackage{amssymb}
\newcommand\MYhyperrefoptions{bookmarks=true,bookmarksnumbered=true,
pdfpagemode={UseOutlines},plainpages=false,pdfpagelabels=true,
colorlinks=true,linkcolor={blue},citecolor={blue},urlcolor={black},
pdftitle={Bare Demo of IEEEtran.cls for Computer Society Journals},
pdfsubject={Typesetting},
pdfauthor={Michael D. Shell},
pdfkeywords={Computer Society, IEEEtran, journal, LaTeX, paper,
             template}}
\ifCLASSINFOpdf
\usepackage[\MYhyperrefoptions,pdftex]{hyperref}
\else
\usepackage[\MYhyperrefoptions,breaklinks=true,dvips]{hyperref}
\usepackage{breakurl}
\fi
\usepackage{array}
\newcommand{\CP}[1]{{\textcolor{red}{[CP: #1]}}}

\begin{document}
%


\newcommand{\MYtitle}{Static and Dynamic Strategies for Influencing Opinions in Social Networks}
\title{\MYtitle}
%
%
%
%

\author{Paolo~Tarantino$^\dagger$,~Fabio~Mazza$^\dagger$,~Carlo~Piccardi,~and Francesco~Pierri$^*$
\thanks{All authors are with the Department
of Electronics, Information, and Bioengineering, Politecnico di Milano, Milan, Italy.\protect\\
$^\dagger$These authors contributed equally to this work.\protect\\
$^*$Corresponding author:  francesco.pierri@polimi.it
}
}

\IEEEtitleabstractindextext{%
\begin{abstract}
The ability of a small set of coordinated actors to manipulate opinions in online social networks poses a serious challenge to the fairness and integrity of public debate. 
We investigate this problem by studying how targeted \textit{stubborn agents} can shift the average opinion of a network governed by the Hegselmann–Krause bounded-confidence dynamics.
Experiments are conducted on weighted LFR benchmark networks with community structure, using multiple node-selection strategies based on degree, strength, PageRank, betweenness, k-coreness, s-coreness, and salience. 
We compare static interventions, in which {stubborn agents} keep a fixed extreme opinion, with dynamic interventions, in which their opinion gradually evolves from moderate to extreme values. 
Results show that dynamic strategies are substantially more effective than static ones, as they exploit bounded-confidence dynamics to progressively recruit intermediate agents and extend influence across the network.
In contrast, static strategies tend to create early opinion separation and therefore have a more limited reach. 
We also find that while some centrality measures offer advantages in static settings, dynamic interventions can achieve strong performance even with simple or random node selection. 
Overall, the study clarifies how intervention design and target selection interact in shaping collective opinions, with implications for understanding and countering manipulation in social networks.
\end{abstract}

}

\maketitle

\IEEEdisplaynontitleabstractindextext

%
\IEEEpeerreviewmaketitle

\IEEEraisesectionheading{\section{Introduction}\label{sec:introduction}}

%
%
%
%
\IEEEPARstart{P}{ublic} opinion has long been shaped by communication systems that amplify selected messages, narratives, and viewpoints.
In contemporary digital environments, this process has become faster, more targeted, and more scalable because online social platforms enable the rapid dissemination of information across large and heterogeneous audiences \cite{vosoughi2018spread}. 
As a result, the same mechanisms that support information access and civic participation can also be exploited to steer collective attention, reinforce biases, and condition opinion formation \cite{boulianne2023role}. 
This dual role has made the study of influence and manipulation in online networks increasingly important, especially in contexts where the fairness and impartiality of public debate are at stake \cite{sambrook2012delivering}. 

The effects of coordinated influence campaigns have been documented in several socially relevant domains, including public health communication during the COVID-19 pandemic \cite{unlu2024unveiling,himelein2021bots,apuke2021fake,van2020inoculating,balakrishnan2022infodemic,patwa2021fighting,rocha2021impact,moscadelli2020fake,caldarelli2021flow,nogara2024misinformation}, electoral and referendum campaigns such as the 2016 and 2020 U.S. elections and Brexit \cite{almond2022trolls2016,o2017russian,badawy2018analyzing,jamieson2020cyberwar,ferrara2020characterizing,khan2020final,holler2021human,burgess2017here,bastos2022botsBrexit}, and online discourse surrounding international conflicts \cite{xu2025social,weaver2017social,marigliano2024analyzing,smart2022istandwithputin,command2024russian}. 
More broadly, growing evidence points to a significant interplay between online influence dynamics and political polarization \cite{gauthier_political_2026}.

To mitigate these forms of opinion manipulation, it is first necessary to understand the mechanisms through which they operate. 
This makes it possible to characterize potential threats, identify the conditions under which they are most effective, and pinpoint vulnerable regions of the underlying social network. 
A large body of research has proposed agent-based models of opinion dynamics \cite{abelson1967mathematical,degroot1974reaching,krause2000discrete,Rainer2002-RAIODA,galam2008sociophysics,friedkin2016network,bolzern2018opinion}. 
More recently, the growing interest in propaganda optimization in political, advertising, and commercial settings \cite{carletti2006make,crain2019political,zeng2014behavior} has motivated the design of targeted influence strategies. 
Many of these approaches seek network configurations that maximize diffusion \cite{peng2020network,nematzadeh2014optimal} or rely on optimization-based metrics that may become computationally demanding on large networks \cite{yildiz2013binary,chiyomaru2022adversarial,ijcai2017p124,kempe2003maximizing,banerjee2020survey}. 
In such cases, simpler and more scalable heuristics based on standard centrality measures may be preferable, as they require only limited structural information. 
However, most existing strategies assume fixed interventions and do not explicitly consider influence policies that evolve over the course of the opinion-dynamics process.

In this work, we compare several centrality-based targeting strategies for influencing collective opinion in a social network. 
To this end, we study the Hegselmann--Krause opinion-dynamics model and examine how the introduction of a fraction of \textit{stubborn agents} affects the network’s average opinion under different initial conditions. 
We consider both a static intervention setting, in which \textit{stubborn agents} keep a fixed opinion, and a dynamic one, in which their opinion changes over time. 
Our goal is to assess which centrality measures provide the most effective and robust criteria for selecting target nodes, and how their performance varies across intervention strategies and simulation settings.

The remainder of the paper is organized as follows. Section~\ref{sec:models} reviews the main agent-based models of opinion dynamics, with particular emphasis on the Hegselmann--Krause model. Section~\ref{sec:strategies} surveys influence strategies proposed in the literature and introduces a classification of intervention types. Section~\ref{sec:materials} describes the network model adopted in our experiments, presents the centrality measures considered in the analysis, and details the simulation protocol and evaluation metrics. Section~\ref{sec:experimentalresults} reports and discusses the results. Finally, Section~\ref{sec:conclusion} concludes the paper and outlines possible directions for future research.


\section{Models of Opinion Dynamics}\label{sec:models}

Individual opinions evolve as a result of personal predispositions and social influence. 
A population can be modeled as a network of agents whose opinions change over time due to two concurrent forces: (i) an intrinsic tendency or prejudice, reflecting each agent’s private information, background, and media consumption, and (ii) social influence arising from interactions via standard communication channels, such as mass media and face-to-face conversations, as well as online social platforms. The \textit{influence intensity} of these platforms can be adjusted through their algorithms.

Here, we assume that interactions occur through a simple, undirected graph $G = (V, E)$, in which multi-edges and self-loops are not permitted. 
In this model, $V$ is a set of nodes representing agents and $E$ is a set of undirected edges representing the social contacts between agents. The literature discusses two broad modeling paradigms:

$\bullet$ \textit{\textbf{Deterministic (differential/difference) models}}: 
Each agent $i$ has a real-valued opinion $x_i(t)\in \mathbb{R}$. 
In classical ``averaging'' models, agents update their opinions at discrete time steps, as in the French–DeGroot model \cite{degroot1974reaching} and the Krause model \cite{krause2000discrete}, or continuously, as in the Abelson model \cite{abelson1967mathematical}. 
They do so by averaging their own opinion and those of their neighbors:
\begin{equation}
    x_i(t+1)=\sum_j w_{ij}x_j(t), \quad \dot{x}_i(t)= -\sum_j l_{ij}x_j(t),
\end{equation}
where $W = [w_{ij}]$ is a row-stochastic ``trust'' matrix and $L=[l_{ij}] = I-W$ is the corresponding Laplacian.
Under mild connectivity conditions, these models converge to a consensus state. Therefore, they cannot explain persistent disagreement among agents \cite{proskurnikov2017tutorial}, which is instead the typical outcome of nonlinear couplings. For example, the Hegselmann-Krause model \cite{Rainer2002-RAIODA} yields clustering of the agents rather than full consensus (see below).

$\bullet$ \textit{\textbf{Stochastic (microscopic) models}}:
These models are often Markov chains with a finite set of opinions. 
Examples include the voter model and its modifications \cite{bolzern2018opinion,bolzern2019opinion,bolzern2020opinion,bolzern2021effect,bolzern2023multi}, Galam’s majority dynamics \cite{galam2008sociophysics}, and the continuous-time ``opinionation'' model \cite{friedkin2016network}, in which each agent’s jump rates depend on their ``prejudice'' and the current mix of neighbors’ opinions. 
These stochastic models capture spontaneous (i.e., endogenous) changes and the probabilistic nature of interpersonal influence. They also allow for the calculation of long-term expected vote shares and notions of individual ``social power'' \cite{bolzern2023manipulating}. 
Stochastic models naturally handle random neighbor sampling, asynchronous events, individual-level volatility, and bounded, discrete opinion sets. While they can predict fluctuations around consensus and the impact of \textit{stubborn agents} (``zealots''),  they typically require more sophisticated probabilistic tools, such as martingales and mean-field approximations, or large-scale simulations.

\subsection{Hegselmann–Krause Model}
\label{sec:hkmodel}

This study adopts the heterogeneous Hegselmann–Krause (HK) model \cite{Rainer2002-RAIODA}, which describes opinion dynamics under \textit{bounded confidence}. 
Each agent $i$ of the population is equipped with a continuous dynamical variable $x_i(t) \in [0, 1]$, representing their opinion, and a fixed quantity $\epsilon_i$, which models their confidence.
Nodes only interact with neighbors whose opinions differ by less than their confidence range, i.e., $x_j\in[x_i-\epsilon_i,x_i+\epsilon_i]$. 
Thus, considering both the network constraint and the confidence range, the set of interactions for agent $i$  is given by:
\begin{equation}
\label{eq:seti}
    I(i,\vec{x})=\left\{1 \leq j \leq N \ | \ (i,j) \in E\ \land |x_i-x_j| \leq \epsilon_i \right\}.
\end{equation}
Opinions are updated synchronously at discrete time steps.
Agent $i$ adopts the average opinion of set $I(i,\vec{x})$:
\begin{equation}
\label{eq:updatehk}
    x_i(t+1)=\frac{1}{|I(i,\vec{x}(t))|}\sum_{j\in I(i,\vec{x}(t))}x_j(t).
\end{equation}
We adapted the update equation \eqref{eq:updatehk} to weighted networks. 
More precisely, the opinions of neighbors contained in $I(i,\vec{x})$ are weighted by $w_{ji}$, i.e., the weight of the edge connecting them to node $i$. 
This gives more importance to neighbors with a strong connection to the updating agent:
\begin{equation}
\label{eq:updatehkweighted}
    x_i(t+1)=\frac{1}{\sigma(i,\vec{x}(t))}\sum_{j\in I(i,\vec{x}(t))} w_{ji} x_j(t).
\end{equation}
Here, $\sigma(i,\vec{x}(t))=\sum_{j\in I(i,\vec{x}(t))} w_{ji}$ is the strength of node $i$ restricted to its neighbors contained in $I$ at time $t$. 
Including this feature is essential for accurately capturing the strength of influence that characterizes real-world networks \cite{barrat2004architecture,newman2004analysis}, and it is particularly important for modeling social media dynamics \cite{bolzern2023manipulating}. 
 
By restricting each agent’s update to the subset of neighbors whose opinions lie within their individual confidence threshold $\epsilon_i$, the HK model naturally generates stable opinion clusters that closely resemble real‐world community polarization \cite{blondel2009krause}. 
Despite its nonlinear confidence bounds, the model is amenable to rigorous convergence analysis via fixed‐point arguments and spectral methods~\cite{bhattacharyya2013convergence}. 
The model is known to converge in finite time to a stable configuration of opinion clusters. 
It has been proven that every trajectory reaches a steady state within a polynomial number of synchronous update steps \cite{blondel2009krause}. 
Subsequent analyses have characterized the cluster‐formation process itself. 
In \cite{lorenz2005stabilization}, it was shown that once opinion distances exceed the confidence bound, agents remain in distinct, non-interacting clusters. 
In \cite{mirtabatabaei2012opinion}, the results were extended to heterogeneous confidence levels. This extension predicted both the number and composition of the final clusters based on the distribution of the individual $\epsilon_i$ values. 

%



\section{Strategies for Influencing Opinions}\label{sec:strategies}

Since the early 2010s, we have witnessed the rapid development of influence strategies within social networks due to the exponential growth of social media and the advent of targeted online marketing. 
These strategies have been used not only for marketing purposes, as in the case of the so-called ``Internet Water Army'' \cite{zeng2014behavior}, but also in many other sectors, including politics \cite{aral2019protecting}. 
Indeed, it has been shown that both the 2016 \cite{badawy2018analyzing,jamieson2020cyberwar} and 2020 \cite{ferrara2020characterizing} US elections underwent significant attempts at manipulation and distortion of the results caused by fake social accounts \cite{khaund2021social}, which inserted themselves into online debates with the aim of spreading propaganda and false news \cite{cheng2020dynamic,mihaylov2018dark}. 
One recurring strategy is to transform nodes in the network into ``stubborn agents'' with a strong bias toward a certain opinion.
Examples of this strategy in opinion‐dynamics models can be seen in \cite{yildiz2013binary}, where agents are inserted into a generalized voter model and influence their neighbors through unidirectional influence links; in \cite{kuhlman2013controlling}, where the highest-degree nodes are targeted; and in \cite{romero2020zealotry}, where two external controllers clash to sway the average opinion toward two opposite extremes. 
An application of this strategy in a Hegselmann-Krause model is presented in \cite{brooks2020model}.
Other research uses the concept of social power as defined in \cite{french1956formal}, as seen in \cite{bolzern2018opinion} and \cite{bolzern2023manipulating}. 
The latter analyzes the effects of attacks on different network communities. This aspect of opinion manipulation has been studied in \cite{noorazar2020recent,weng2013virality,romero2013interplay,nematzadeh2014optimal,hu2014effects,peng2020network,tang2021susceptible,lee2025cultural}, which have produced diverse results depending on the properties of the networks and the diffusion models used. 
In \cite{chiyomaru2022adversarial}, small perturbations are made to the edge weights within a voter model to alter the voting dynamics. 
Further strategies for majority-based models can be found in \cite{ijcai2017p124,zehmakan2021majority}, and a summary of manipulation strategies applied to deterministic and stochastic models can be found in \cite{noorazar2020recent}. 
Finally, some research has focused on maximizing the effects of these attacks, as seen in \cite{kempe2003maximizing,banerjee2020survey,han2020clusters}.

In opinion‐dynamics models, influence strategies can be  classified as either \textit{soft} or \textit{hard attacks}, depending on whether they subtly bias interaction rules or directly impose fixed opinions through dedicated agents \cite{bolzern2023manipulating}.

\subsection{Soft Attacks}
Soft attacks indirectly modify the dynamics to tilt the ``playing field'' in favor of a desired outcome. 
The objective is to shift the average standalone opinion of certain agents \cite{bolzern2023manipulating,noorazar2020recent}. 
Common approaches include the following:
\begin{itemize}
    \item Tuning interaction parameters, such as the confidence bound $\epsilon$ in bounded‐confidence models, can be achieved through manipulative advertising \cite{crain2019political} or monetary incentives \cite{zeng2014behavior}.
    \item Adjusting the weights $w_{ij}$ in DeGroot‐style averaging increases or decreases receptivity to particular viewpoints~\cite{forster2016trust}.  
    \item Broadcasting external fields similar to mass‐media opinion sources can gradually shift all agents’ private biases over time. This can be done by adjusting the probability of interacting with a ``media'' node \cite{pineda2015mass,carletti2006make}.
    \item Altering the network topology by adjusting  edge weights or rewiring can amplify connectivity around pro‐target clusters or isolate dissenting groups, indirectly steering the consensus \cite{chiyomaru2022adversarial,ninomiya2025mitigating}.
\end{itemize}
These interventions are typically  smooth and widespread. 
Although agents have complete freedom to update, the landscape of influences favors the attacker’s goals.
It is challenging to detect these subtle biases in real networks, and effectively implementing them often requires extensive global knowledge or coordination.

\subsection{Hard Attacks}
In contrast, hard attacks introduce one or more stubborn (or ``zealot'') agents into the social network. 
These agents have fixed opinions that do not change over time or under the influence of their neighbors' opinions  \cite{khaund2021social,cheng2020dynamic}. 
\begin{itemize}
    \item In stochastic models, \textit{stubborn agents} influence the random‐walk dynamics of opinions, causing them to converge toward the agent's own value \cite{mukhopadhyay2020opinion,mobilia2007role}.
    \item In deterministic averaging frameworks, such as the DeGroot model \cite{degroot1974reaching}, \textit{stubborn agents} can be included in every neighbor‐average update as immovable boundary conditions \cite{golub2010naive}. 
    In contrast,  the Hegselmann–Krause model \cite{Rainer2002-RAIODA} states that each non-stubborn agent includes a stubborn node in their update only if their confidence is sufficiently large. Thus, the influence of a stubborn agent on the population is filtered by the confidence threshold; only agents ``close enough'' in opinion will be pulled toward the zealot value.
\end{itemize}



\section{Materials and Methods}\label{sec:materials}

\subsection{Network Model}
\label{subsec:lfrbenchmark}

In this work, we implement the HK process on graphs defined by the Lancichinetti–Fortunato–Radicchi (LFR) benchmark model \cite{lancichinetti2009benchmarks}, which is an algorithm that generates networks with power-law distributions for both node degree and edge weight, as well as a predefined community structure. These features allow the network to closely resemble real-world networks, making them essential for modeling phenomena such as intra-community consensus, inter-community polarization, and ``echo-chamber'' effects in opinion dynamics \cite{leicht2008community}.  

We used the implementation provided by the authors \cite{lfrnetcode} with the following parameters to construct the desired networks: $N=1000$ or $2000$ nodes, average degree $\langle k \rangle=20$, maximum degree $k_{\max}=200$ , mixing parameter for strength $\mu_w=0.1$, minimum and maximum community size $c_{\min}=20$ and $c_{\max}=50$. 
To build a weighted network, the algorithm first creates an unweighted network and then assigns a positive real weight to each link. 
This assignment is based on two parameters, $\beta$ and $\mu_w$. The first parameter is set to $\beta=1.5$ by default and assigns a strength $\sigma_i$ to each node according to $\sigma_i=k_i^{\beta}$, which mimics the strength-degree power-law relation frequently observed in real-world weighted networks. 
The second parameter, $\mu_w$, is then used to assign the internal strength $\sigma_i^{(in)}$, which is the strength directed inside the community of node $i$, as follows $\sigma_i^{(in)}=(1-\mu_w)\sigma_i$. Using this procedure, we generated 20 network instances for each network size $N=1000$ and $2000$.

\begin{figure}[!t]
    \centering
    \includegraphics[width=\linewidth]{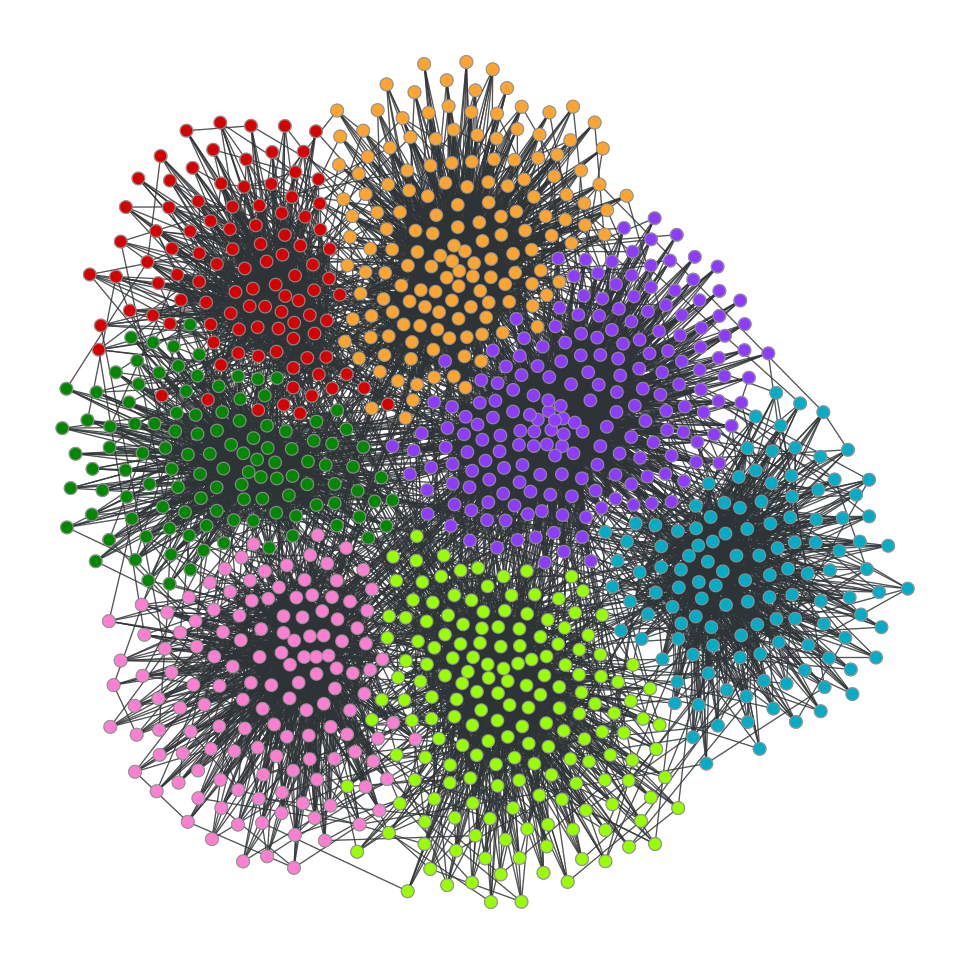}
    \includegraphics[width=\linewidth]{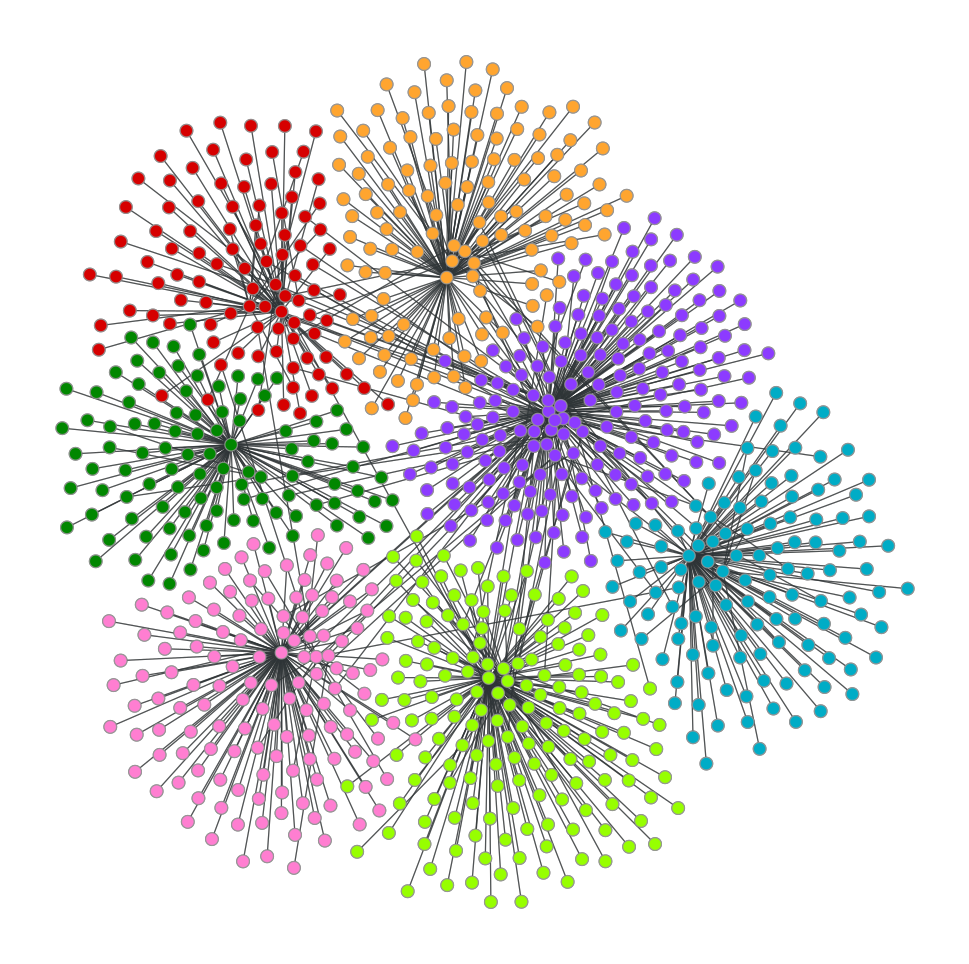}
    \caption{Top panel: An LFR network generated as described in Sec. \ref{subsec:lfrbenchmark} ($N=1000$): colors correspond to communities. Bottom panel: the High-Salience Skeleton (see Sec. \ref{sec:centralities}) of the network in the top panel.}
    \label{fig:lfrnetwork}
\end{figure}

One of the network instances is shown in Fig. \ref{fig:lfrnetwork} (top panel), together with the corresponding High-Salience Skeleton (bottom panel), which is a backbone containing almost all of the nodes (94\%) but only less than 9\% of the edges (see Sec. \ref{sec:centralities} for definition and details).

\subsection{Network Centralities}
\label{sec:centralities}

To characterize the structural roles of the nodes in our network and guide the selection of \textit{stubborn agents}, we compute several centrality measures \cite{sabidussi1966centrality,freeman1977set,bonacich1987power,borgatti2006graph,opsahl2010node,seidman1983network,page1999pagerank}. Each metric captures a different aspect of ``importance'' within the graph. By comparing these different centralities, we can evaluate the impact of connectivity-based (e.g., degree or coreness), path‐based (e.g., betweenness, salience), and recursive (e.g., PageRank) notions of importance on the ability of a small set of agents to influence overall opinion formation.

$\bullet$ \textit{\textbf{Degree Centrality}}: It corresponds to the number of neighbors of node $i$:
\begin{equation}
    d_i = \sum_{j=1}^N a_{ij},
\end{equation}
where $A=[a_{ij}]$ is the \textit{adjacency matrix}, which has $a_{ij} = a_{ji} = 1$ if an edge connects nodes $(i,j)$, $a_{ij} = a_{ji} = 0$ otherwise ($a_{ii} = 0$ $\forall i$ because self-loops are not allowed). 
Nodes with a high degree have many direct connections, enabling them to communicate quickly with a large number of neighboring agents.

$\bullet$ \textit{\textbf{Betweenness Centrality}}: 
It measures the fraction of the shortest paths that include node $i$:
\begin{equation}
    b_i = \sum^N_{\substack{j,k=1 \\ j\neq i \neq k}}  \frac{\eta_{jk}(i)}{\eta_{jk}}.
\end{equation}
Here, $\eta_{jk}$ is the total number of shortest paths from $j$ to $k$ and $\eta_{jk}(i)$ is the number of those paths that pass through $i$. 
Nodes with high betweenness act like bridges and can control the flow of information across different regions of the network.

$\bullet$ \textit{\textbf{Strength}}: In a weighted network with weight matrix $W=[w_{ij}]$, the strength of node $i$ is defined as follows:
\begin{equation}
    \sigma_i = \sum_{j=1}^N w_{ij},
\end{equation}
where $w_{ij}$ is the weight of the edge $(i,j)$. 
This measures the overall intensity of node $i$'s connections.

$\bullet$ \textit{\textbf{PageRank}}: It was originally developed for web page ranking and assigns each node $i$ the following score:
\begin{equation}
    \pi_i = \gamma \sum_{j=1}^N \pi_j \frac{w_{ji}}{\sigma_j} + (1- \gamma) \frac{1}{N},
\end{equation}
where $N$ is the number of nodes in the graph, $\sigma_j$ is the strength of node $j$, and $\gamma \in [0,1]$ is a damping factor. 
Nodes linked to other well-connected nodes receive a higher PageRank, reflecting both local and global influence. 
In our experiments, we set the damping factor to the standard value $\gamma = 0.85$.

$\bullet$ \textit{\textbf{k-coreness}}: The $k$-core of a graph is the largest subgraph in which every node has an internal degree of at least $k$. 
The coreness of a node $i$ is the largest $k$ such that $i$ belongs to the $k$-core. 
Nodes with high coreness are deeply embedded in the core of the network.

$\bullet$ \textit{\textbf{s-coreness}}: It is analogous to $k$-coreness, but it is based on node strength rather than degree.
The $s$-core of a weighted graph is the largest subgraph in which every node has an internal strength of at least $s$.
The $s$-coreness of a node $i$ is the largest value $s$ such that $i$ belongs to the $s$-core.
This measure captures resilience under weighted node removal: nodes with high $s$-coreness are deeply embedded in the weighted core of the network.

$\bullet$ \textit{\textbf{Salience}}: Edge salience categorizes edges based on their intrinsic properties within the network. It can be seen as the shared consensus among nodes about the importance of an edge \cite{grady2012robust}. It revolves around the notion of the average Shortest Path Tree (SPT):
\begin{equation}
    S = \langle T \rangle =\frac{1}{N}\sum_{r=1}^N T(r),
\end{equation}
where, given a reference node $r$, $T(r)$ is the symmetric $N\times N$ matrix summarizing the shortest paths from $r$ to all other nodes ($t_{ij}=1$ if the edge $(i,j)$ is part of at least one shortest path, $t_{ij}=0$ otherwise). $S$ is the superimposition of all SPTs, so that $0 \leq s_{ij} \leq 1$ is a consensus variable defined by the ensemble of nodes that quantifies the fraction of SPTs in which the edge $(i,j)$ participates. If $s_{ij}=1$, then the edge $(i,j)$ is essential for all nodes. If $s_{ij}=0$, then the edge has no role. If, for example, $s_{ij}=0.5$, then the edge is important for only half of the nodes. Real-world networks are often scale-free \cite{barabasi1999emergence}, with a few hubs and many weakly connected nodes. 
If they are weighted, they usually exhibit power-law distributions for both node degrees and edge weights. 
In such networks, it is possible to obtain a robust classification of edges based on edge salience because the latter typically has a bimodal distribution on the unit interval, accumulating at the extremes~\cite{grady2012robust}.
Taking only the edges with $s_{ij} \approx 1$, we define the High-Salience Skeleton (HSS), a robust, disassortative backbone with a scale-free topology that is often divided into multiple components.
The concept of salience can then easily be extended from edges to nodes. 
We define the salience of node $i$ as the sum of the salience values of all its incident edges:
\begin{equation}
    s_i = \sum_{j=1}^N s_{ij}.
\end{equation}
Note that, in the limit case where $s_{ij}=1$ for each edge $(i,j)$ in the HSS, and $0$ otherwise, $s_i$ reduces exactly to the HSS degree of $i$, i.e., the number of edges incident on $i$ in the HSS.

\subsection{Driving Opinions: Static and Dynamic Strategies}
\label{subsec:strategies}

\begin{figure}[!t]
    \centering
    \includegraphics[width=\linewidth]{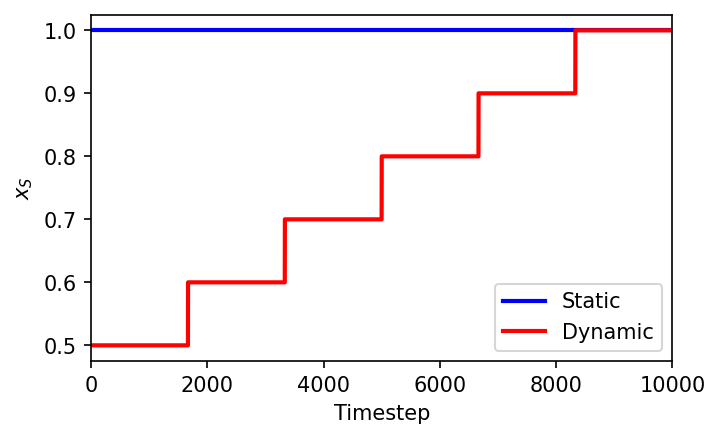}
    \caption{The temporal patterns of the opinion of \textit{stubborn agents} under static and dynamic strategies with final $x_S=1$.}
    \label{fig:visualizationofstrategies}
\end{figure}

In our experiments, we select a small fraction of nodes and transform them into \textit{stubborn agents}, which are biased toward an extreme opinion and can shift the average opinion toward their own extreme value. 
We set the target at  the upper extreme of the possible opinion range, which is 1. 
The \textit{stubborn agents}' opinions are fixed by assumption, so they are not influenced by their neighbors. 
Thus, their opinion value is not updated according to the HK dynamic model.
Stubborn nodes can be chosen randomly or according to a centrality-based logic that selects the top-ranking fraction of agents based on a specific centrality metric. 
All \textit{stubborn agents} have the same opinion value $x_{S}$, which is set using two  strategies: \textit{static} and \textit{dynamic} (Fig. \ref{fig:visualizationofstrategies}). 

$\bullet$ \textit{\textbf{Static strategy}}: The value of all \textit{stubborn agents}' opinions is set to $x_{S}=1$ for the entire simulation horizon (see Fig. \ref{fig:visualizationofstrategies}, blue curve). 
We also experimented with $x_{S}=0$ and obtained similar results because the setting is symmetrical. We omit these results for brevity's sake.

$\bullet$ \textit{\textbf{Dynamic strategy}}: 
The stubborn opinion $x_{S}$ increases over time within the range $x_{S}\in [0.5,1]$.
More precisely, the simulation is divided into six equal periods, and $x_{S}$  increases by $0.1$ after each period.
This process begins with $x_S=0.5$ and ends with $x_S=1$ (see Fig. \ref{fig:visualizationofstrategies}, red curve). 
This approach addresses the issue of  rapid conditioning in the HK process, which excludes nodes with  distant opinions or small confidence values. 

\subsection{Initialization of the HK Model}
\label{subsec:hkmodelinit}
For each simulation, the initial opinions and confidence values are drawn from uniform distributions. That is, $x_i(0) \in \mathcal{U}[0,1]$ and $\epsilon_i\in \mathcal{U}[\epsilon_l,\epsilon_u]$, $\forall i = 1,...,N$, with $\epsilon_l, \epsilon_u \in [0,1]$ and $\epsilon_l \leq \epsilon_u$. 

In order to accurately represent the interactions of a real-world network,  the range of possible confidence values must be limited.
The lower threshold has been experimentally set to $\epsilon_l = 0.05$: nodes close to this lower bound are unlikely to change their opinion yet they still interact with neighbors whose opinions are nearly identical to their own. 
When the upper threshold $\epsilon_u$ is sufficiently large,  the system tends to converge toward global consensus. However, when it is low,  the system fragments into multiple opinion groups, leading to polarization. 
In a real-world context, it is unrealistic for a node to immediately switch from one extreme opinion to another due to the strong influence of a node with an almost opposite view. 
If such a transition occurs, it is gradual and takes time.
Furthermore, global consensus is rarely observed. Typically, one finds a small number -- greater than one -- of large clusters with a broad distribution of opinions. 
For these reasons, the upper threshold has been  experimentally set to $\epsilon_u = 0.25$, which usually results in two or three main clusters of opinions in an uncontrolled process.

\subsection{Simulation Protocol}
\label{subsec:simprotocol}

We carried out two sets of simulations, one for each strategy (static and dynamic). 
In both cases, we performed 50 simulations with random initialization (see Sec. \ref{subsec:hkmodelinit}) on each of the 20 network instances, for all centrality metrics and with a proportion of \textit{stubborn agents} $f_S$ taking the values 0.1\%, 0.2\%, 0.5\%, 1\%, and 2\%. Note that 0.1\% in a network with 1000 nodes corresponds to only one stubborn agent. 

Each simulation runs for $10000$ time steps or stops earlier if the opinions converge and reach a steady state. 
The convergence criterion is as follows:
\begin{equation}
    \label{eq:convergence}
    \sum_{i=1}^N|x_i(t-1)-x_i(t)|<10^{-4}.
\end{equation}
For the dynamic strategy, this criterion is only checked when the \textit{stubborn agents} reach the final opinion value, $x_S=1$.


\section{Results}
\label{sec:experimentalresults}
In this section, we present and discuss the results for a network size of $N=1000$. The results for $N=2000$ are qualitatively similar and available in the Supplementary Material. 
Unless otherwise specified, all results for a given parameter setting are averaged over 50 simulations and 20 network instances.
Confidence bounds are specified whenever relevant. 

\subsection{Static Strategy}

\begin{figure}[!t]
    \centering
    \includegraphics[width=\linewidth]{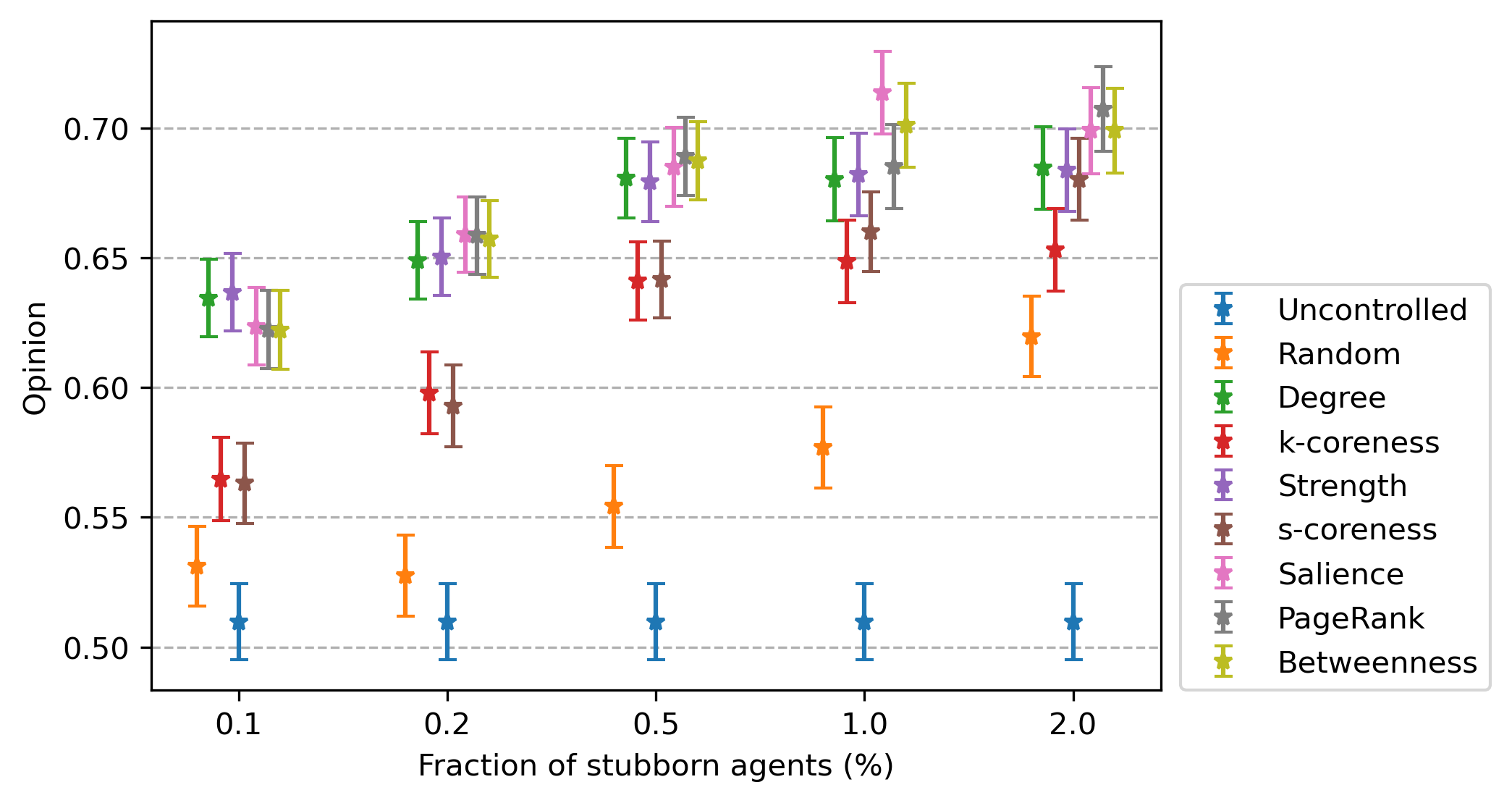}
    \caption{Final average opinion as a function of the fraction $f_S$ of stubborn agents, with fixed opinion $x_S=1$, for the different centrality measures used to select them in the \textit{Static Strategy} setting ($N=1000$).}
    \label{fig:avgbyfrac_opi_1}
\end{figure}

Using the \textit{Static Strategy}, we examine the final average opinion of the population, which is expected to be around $0.5$ in the \textit{Uncontrolled} case. 
Figure \ref{fig:avgbyfrac_opi_1} shows that the increase yielded across various centralities is similar, with the exception of \textit{k-coreness} and \textit{s-coreness} as well as the \textit{Random} baseline. 
Other centralities, such as \textit{Salience}, \textit{Betweenness}, and \textit{PageRank}, outperform the others in most cases. 
This advantage is more evident at small fractions of stubborn agents, where the increase in the final average opinion is about 10 percentage points (roughly from 0.55 to 0.65), whereas it becomes less pronounced at larger fractions, dropping to about 5 percentage points (roughly from 0.65 to 0.7). 
Furthermore, the performance of all centralities increases monotonically with the fraction of stubborn agents, although this effect tends to saturate around $f_S=1\%$.

\begin{figure}[!t]
    \centering
    \includegraphics[width=1.\linewidth]{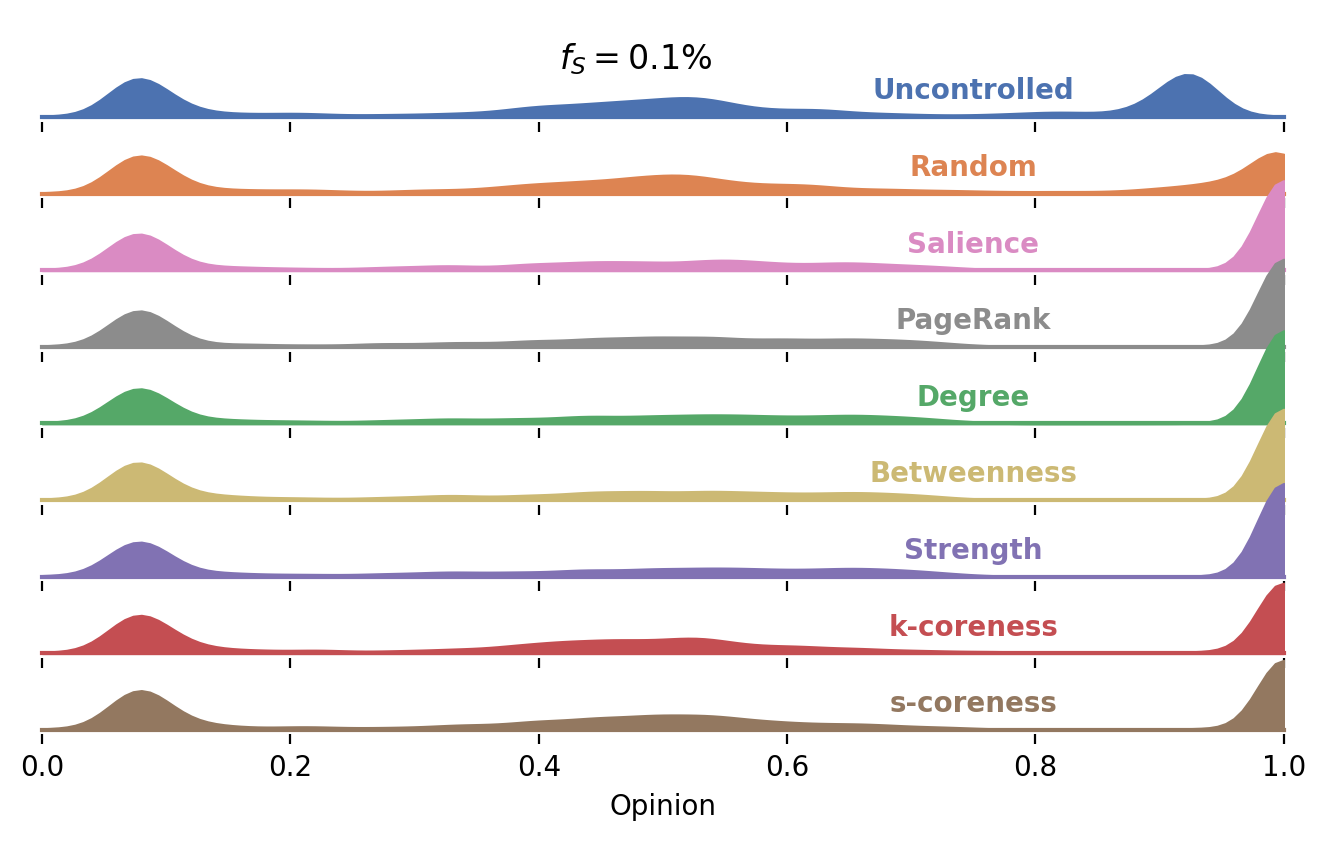}
    \includegraphics[width=1.\linewidth]{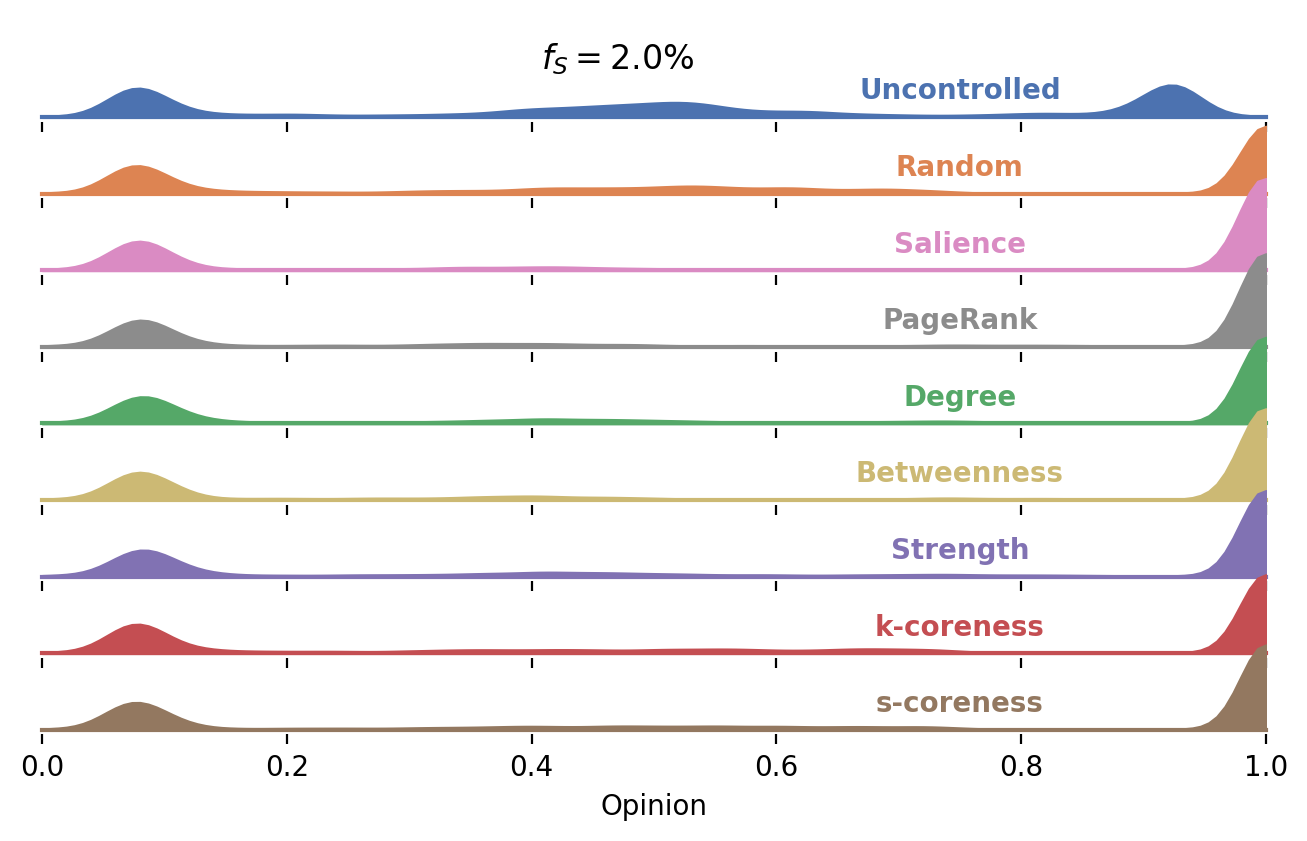}
    
    \caption{Final opinion distribution for two values of the fraction $f_S$ of stubborn agents (top: 0.1\%, bottom: 2.0\%), with fixed opinion $x_S=1$, obtained for all the centrality measures used to select them in the \textit{Static Strategy} ($N=1000$).}
    \label{fig:ridge_opi_1}
\end{figure}

Examining the distribution of opinions at the final time step (Fig. \ref{fig:ridge_opi_1}), we observe that opinions split into two groups: one centered around $x_S=1$ and the other close to zero. 
Comparing these results with the \textit{Uncontrolled} case shows that \textit{stubborn agents} attract only the portion of the population already close to $1$, rather than the entire network. 
Indeed, the part of the distribution far from $x_S=1$ remains nearly unchanged across all centralities and in the \textit{Uncontrolled} case.
A small but noticeable fraction of agents remains at an intermediate opinion around $0.5$; however, this residual group progressively vanishes as $f_S$ increases (see $f_S=2.0\%$).

\begin{figure}[!t]
    \centering
    \includegraphics[width=1.\linewidth]{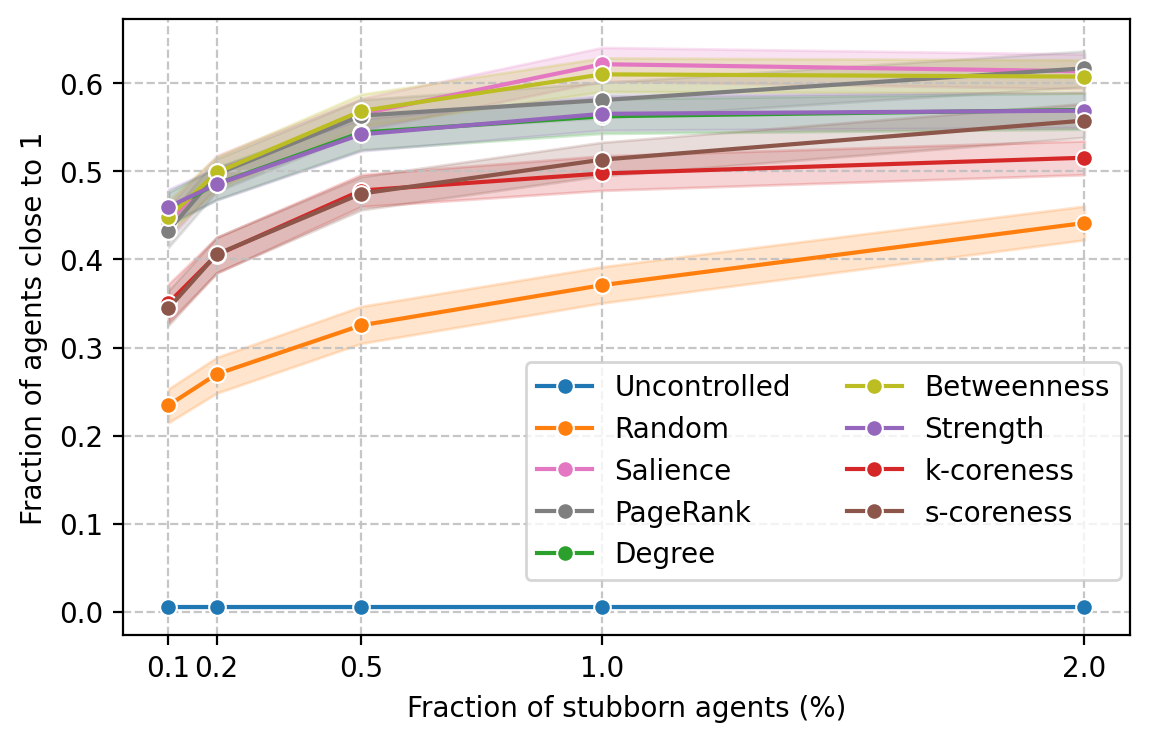}
    \caption{Fraction of the population with final opinion at distance less than $0.05$ from $x_S = 1$, as a function of the fraction $f_S$ of stubborn agents and of the centrality used to select them in the \textit{Static strategy} ($N=1000$).}
    \label{fig:frac_near_opi_1}
\end{figure}

The above observation suggests measuring the number of agents ``captured'' by \textit{stubborn agents}---i.e., those whose final opinions lie close to 1.  
Figure~\ref{fig:frac_near_opi_1} shows the fraction of the nodes $i$ satisfying $ |x_i - x_S| < 0.05$ at the final time.
Among the centrality measures, those based on shortest paths---namely \textit{Salience} and \textit{Betweenness}---together with \textit{PageRank}, yield the best overall performance.  
\textit{Strength} and \textit{Degree} perform comparably well, while \textit{k-coreness} and \textit{s-coreness} generally perform worse.  
We observe again a clear saturation limit in the fraction of the population that can be driven to $x_S = 1$.  
Increasing the fraction of stubborn agents from $f_S = 1\%$ to $f_S = 2\%$ does not significantly improve the performance of the best-performing centrality measures (e.g., \textit{Betweenness} and \textit{Salience}).

\begin{figure}[!t]
    \centering
    \includegraphics[width=1.\linewidth]{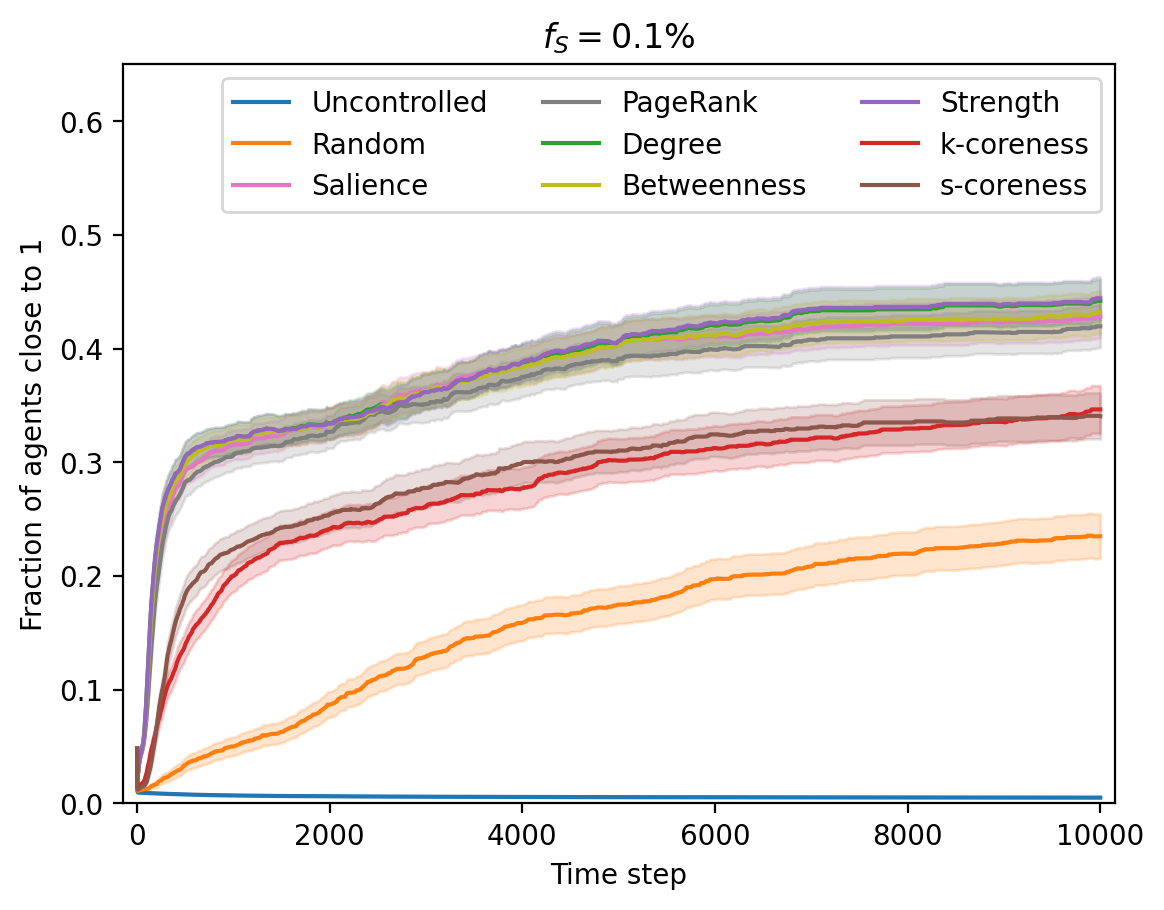}
    \includegraphics[width=1.\linewidth]{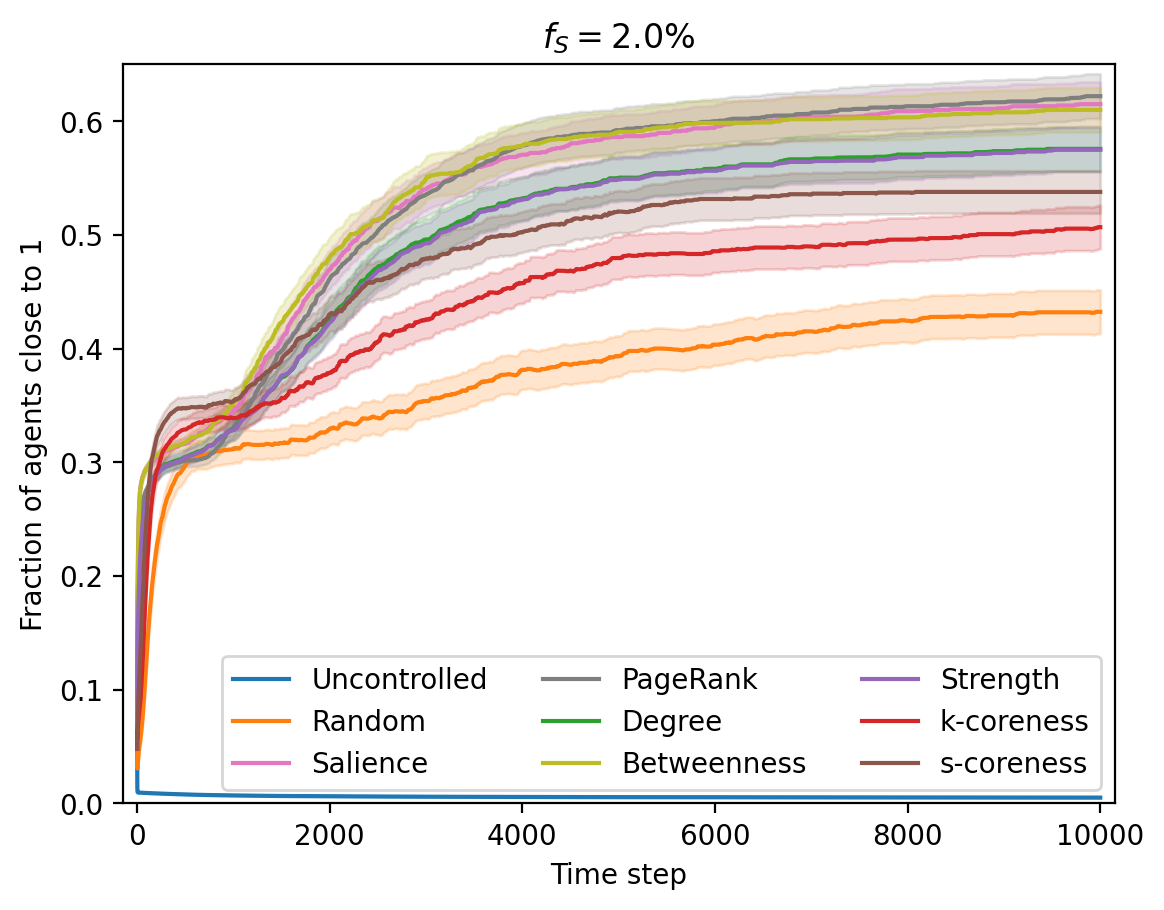}
    \caption{Time evolution of the fraction of the population with opinion at distance less than $0.05$ from $x_S = 1$, for two values of the fraction $f_S$ of stubborn agents (top: 0.1\%, bottom: 2.0\%), with fixed opinion $x_S=1$, obtained for all the centrality measures used to select them in the \textit{Static Strategy} ($N=1000$).}
    \label{fig:frac_near_hist_opi_1}
\end{figure}

To complete the above analysis, Fig.~\ref{fig:frac_near_hist_opi_1} shows the fraction of the population close to the \textit{stubborn agents} over time, for the two extreme values of the fraction of stubborn agents. 
In both cases, a steep initial increase is observed. For the smallest fraction of stubborn agents ($f_S = 0.1\%$, top panel), the behavior is sensitive to the chosen centrality measure from the very beginning. 
In contrast, for the largest fraction ($f_S = 2\%$, bottom panel), the initial steep increase is consistent across all centrality measures, after which an additional phase of increase dependent on the chosen measure follows. 
Overall, within the initial time steps, the stubborn nodes attract approximately about 30\% of the entire population, demonstrating that the static strategy quickly creates a significant opinion divide in the network.

\subsection{Dynamic Strategy}

\begin{figure}[!t]
    \centering
    \includegraphics[width=1.\linewidth]{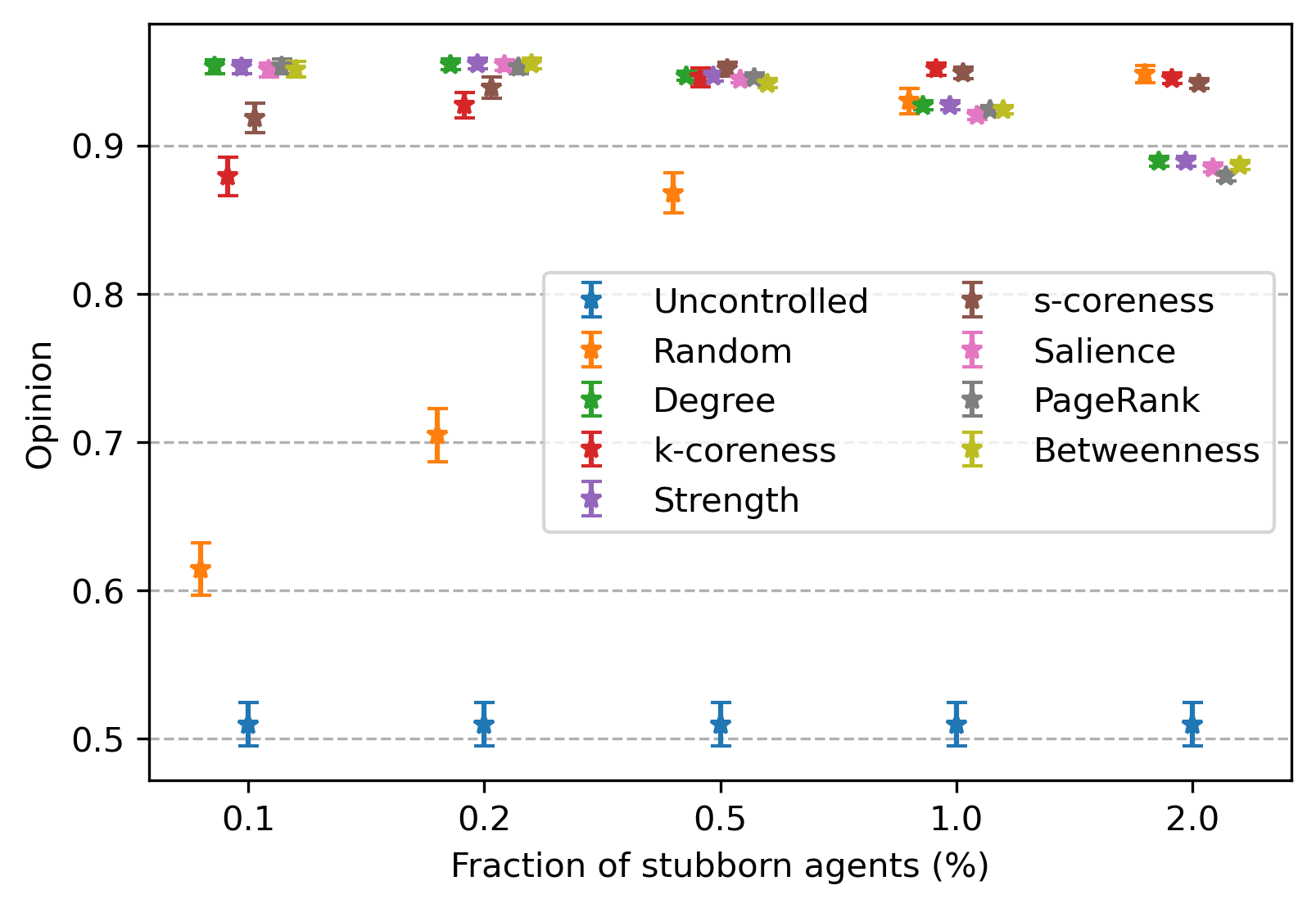}
    \caption{Final average opinion as a function of the fraction $f_S$ of stubborn agents (with final opinion $x_S = 1$) and of the centrality used to select them in the \textit{Dynamic Strategy} ($N=1000$).}
    \label{fig:dynamic_opi_avg}
\end{figure}

\begin{figure}[!t]
    \centering
    \includegraphics[width=1.\linewidth]{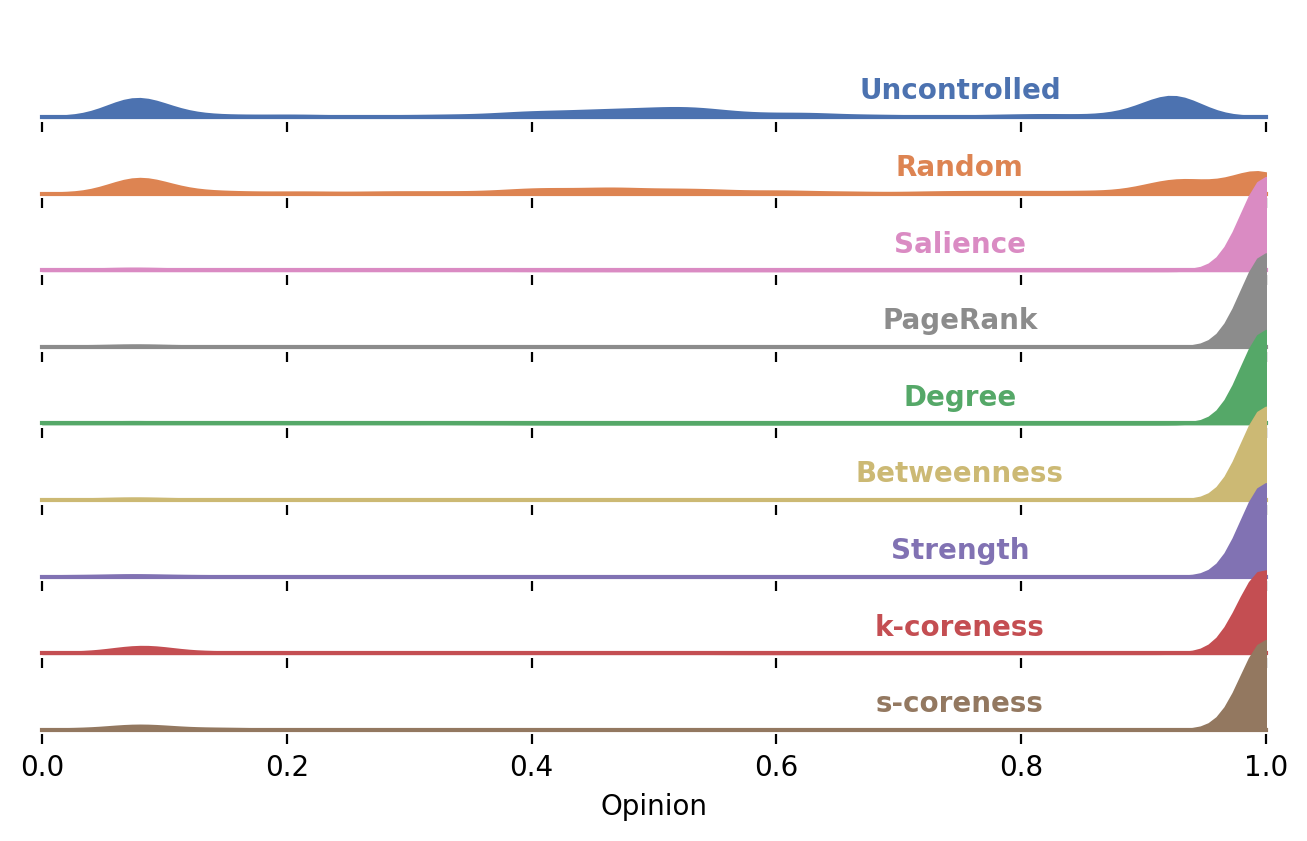}
    \caption{Final opinion distribution for $f_S = 0.1\%$ of stubborn agents, with final opinion $x_S = 1$, obtained for all the centrality measures used to select them in the \textit{Dynamic Strategy} ($N=1000$).}
    \label{fig:ridge_opi_dyn}
\end{figure}

We next present the results of applying the \textit{Dynamic Strategy}, in which the initial opinion of the \textit{stubborn agents} is set to $x_S = 0.5$ and then gradually increases to $x_S = 1$ (see Fig.~\ref{fig:visualizationofstrategies}). 
This procedure attracts a substantially larger fraction of the population, as evidenced by the high final average opinion values in Fig.~\ref{fig:dynamic_opi_avg} and the opinion distributions in Fig.~\ref{fig:ridge_opi_dyn}, which demonstrate the superior effectiveness of this strategy in guiding agents toward the target (compare these distributions with those in Fig.~\ref{fig:ridge_opi_1}).

\begin{figure}
    \centering
    \includegraphics[width=1.\linewidth]{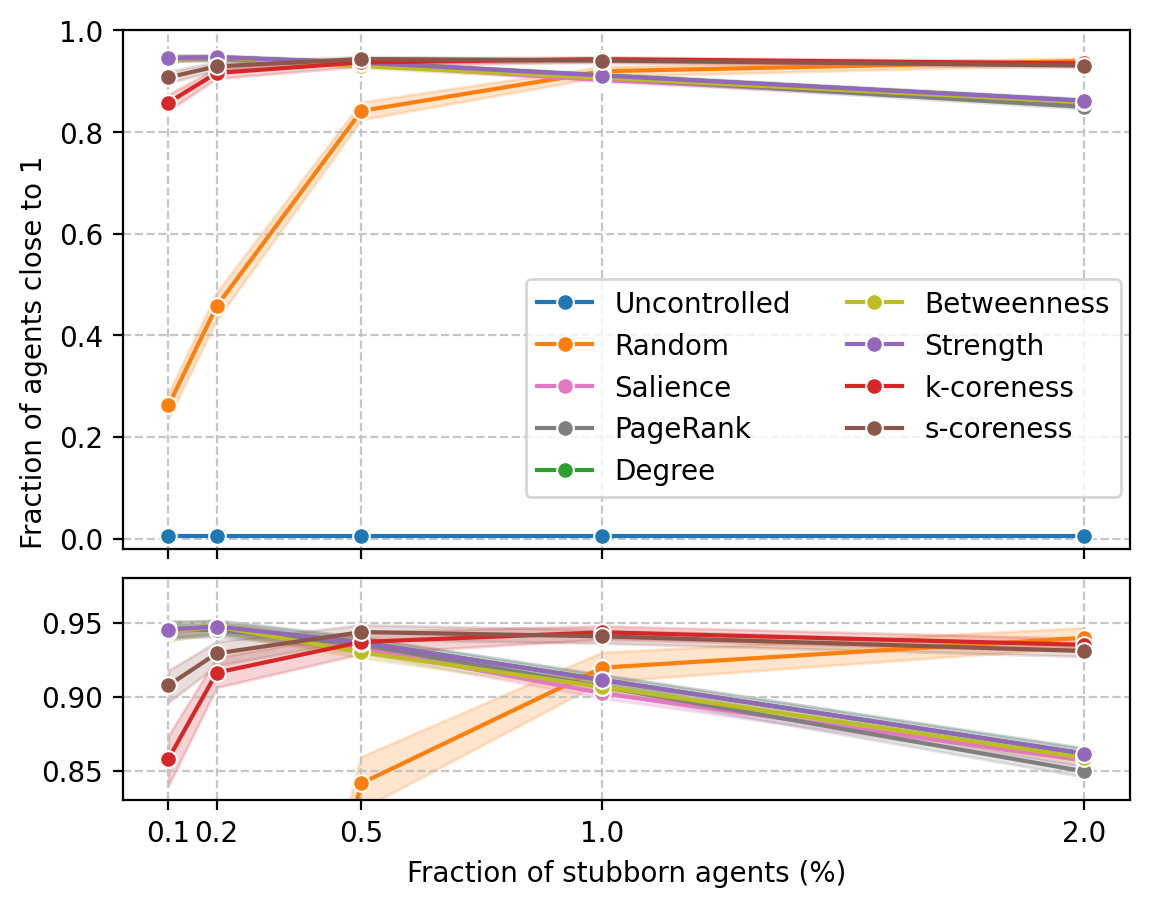}
    \caption{Fraction of the population with final opinion at distance less than $0.05$ from $x_S = 1$, as a function of the fraction $f_S$ of stubborn agents and of the centrality used to select them in the \textit{Dynamic Strategy} ($N=1000$). The bottom panel shows a magnified view of the saturation region from the top panel.}
\label{fig:dynamic_frac_near_opi_1}
\end{figure}

In fact, even a very small fraction of stubborn agents, $f_S = 0.1\%$ (i.e., one agent for $N = 1000$), can drive at least $85\%$ of the population toward the final opinion $x_S = 1$, as shown in Fig.~\ref{fig:dynamic_frac_near_opi_1}. 
However, as the fraction of stubborn agents increases, \textit{k-coreness} and \textit{s-coreness}, together with the \textit{Random} baseline, attract larger fractions of the population, whereas the performance of other centralities decreases. 

\begin{figure*}
    \centering
    \includegraphics[width=0.85\linewidth]{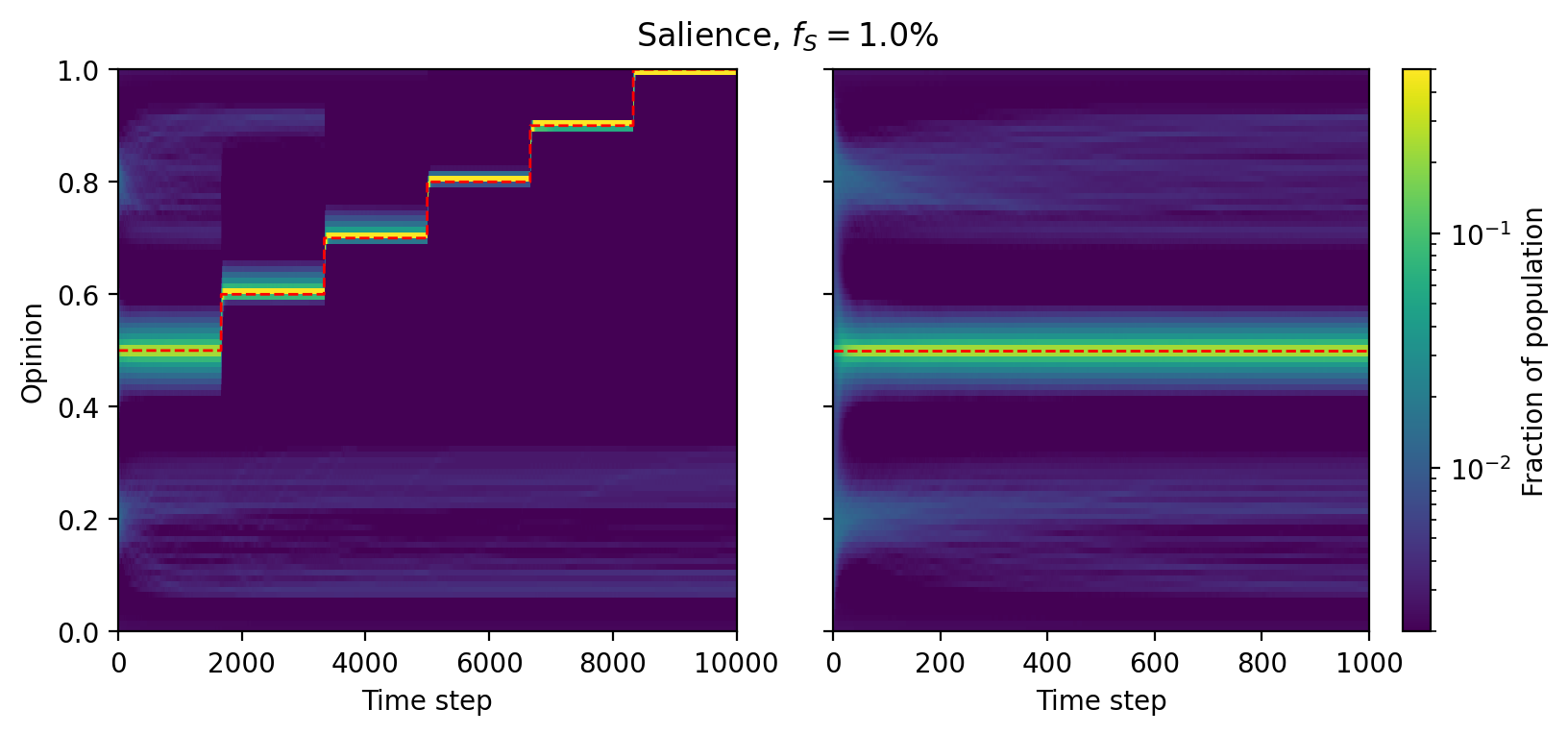}
    \includegraphics[width=0.85\linewidth]{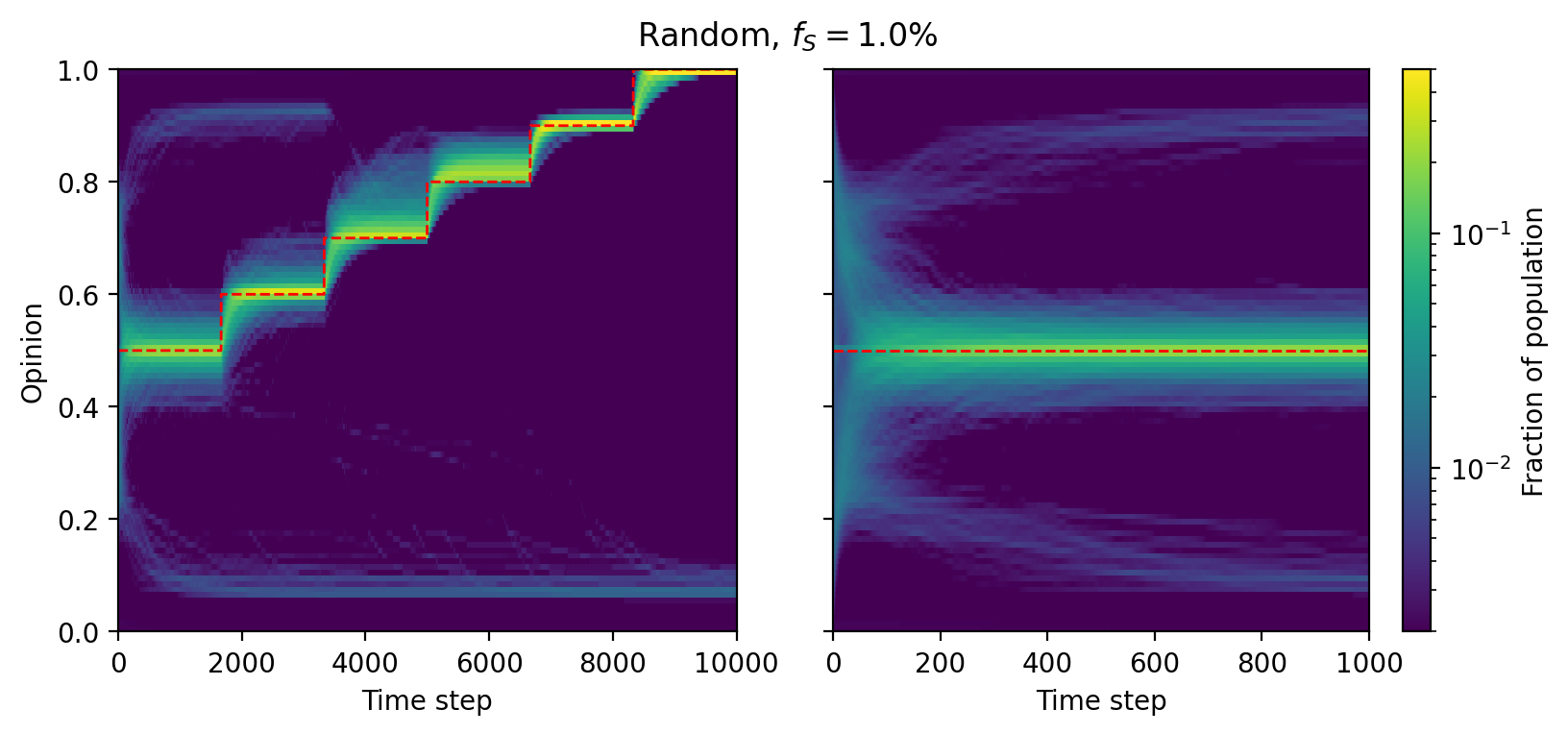}
    \caption{Time evolution of the density of opinions, averaged over 50 simulations and 20 network instances, comparing the dynamics obtained with \textit{Salience} (top panels) and \textit{Random} (bottom panels) for a large fraction of stubborn agents ($f_S = 1\%$). The left panels show the entire simulation. The right panels show only the initial stages up to $t = 1000$. The red dashed lines indicate the opinion of the stubborn agents (\textit{Dynamic Strategy}, $N=1000$).}
    \label{fig:heatmaps_opis_dynamic}
\end{figure*}

This counterintuitive result stems from the underlying opinion dynamics. 
When many nodes are selected using highly effective centralities such as \textit{Betweenness}, \textit{PageRank}, or \textit{Salience}, a significant portion of the population quickly converges to the initial stubborn opinion $x_S = 0.5$ (see the top panels of Fig.~\ref{fig:heatmaps_opis_dynamic}), thereby isolating itself from opinions farther away due to the bounded confidence mechanism of the Hegselmann--Krause (HK) model. 
In contrast, when nodes are chosen randomly or via \textit{k-coreness}/\textit{s-coreness}, the initial convergence to $x_S = 0.5$ is slower or less cohesive, preventing premature lock-in. 
This allows a ``herd effect'' to take place: non-stubborn nodes moving toward the gradually increasing $x_S$ attract additional non-stubborn nodes in a cascading manner. 
Consequently, more agents are drawn to the final opinion $x_S = 1$. 
The herd effect also explains why the dynamic strategy performs remarkably well even with a single well-chosen stubborn node ($f_S = 0.1\%$): a single influential node can trigger a cascade without causing early mass convergence to $x_S = 0.5$.

\subsection{Discussion}
\label{sec:discussion}
It is useful to split the analysis of the results into two parts: the first covers cases with $0.1\%$ to $1\%$ stubborn agents under both static and dynamic strategies; the second examines the $2\%$ case, in light of the peculiar results obtained for that fraction.

Starting with the 0.1\% to 1\% cases, the dynamic strategies consistently outperform their static counterparts. 
Static strategies create a sharp, rapid separation from the rest of the network: due to the bounded confidence mechanism, links to nodes with more distant opinions are completely severed, leading to several well‑defined, isolated clusters. 
In contrast, dynamic strategies gradually attract the network’s nodes over a larger number of time steps. 
This allows nodes with intermediate opinions to interact with the stubborn agents during the initial phase (when the stubborn agents’ opinion is $0.5$), and enables nodes with opinions close to zero to follow the trajectories of those in the intermediate zones. 
The result is a global attraction process in which each agent’s opinion $x_i$ is slowly shifted toward $1$ by neighbors whose values are only slightly higher.

This outcome aligns with real‑world social network dynamics. 
When attempting to shift the average opinion of a group via bots, one rarely starts by posting overtly extremist content. 
Instead, one enters more moderate discussion contexts to attract agents with higher confidence, hoping they will gradually influence users with lower confidence or those at the opposite extreme of the opinion spectrum \cite{lim2022opinion,liang2024silence,luo2023rise,pescetelli2022bots,ninomiya2025mitigating}. 
The theory of mass ideology is strongly supported by the opinion‑dynamics model we adopt \cite{friedkin2011social,friedkin1998structural,noelle1974spiral}, since an agent’s ability to change opinion is proportional to the number of neighbors influencing it, provided the confidence bounds are sufficiently large. 
This explains why a dynamic, gradual conditioning is more effective in most simulations \cite{liang2024silence,friedkin2001norm}.

However, these advantages come at a cost. 
Programming, deploying, managing, and controlling stubborn agents whose opinions adapt dynamically is far more expensive than using static bots that produce the same type of content without communicating with nodes holding very different opinions. 
Moreover, the dynamic process requires significantly more time. 
A slower convergence is more effective than a rapid one if sufficient time is available. 
Conversely, a static strategy typically converges faster, making it preferable when time is limited. 
As expected, a larger fraction of stubborn agents generally yields better results. 
However, introducing a large number of stubborn agents into real‑world networks is not always feasible. 
Therefore, a balance must be struck between the achieved results and the cost of the attack.

For $f_S = 2\%$, the two strategies that performed worst in the previous scenarios now achieve the highest average opinions. 
More effective centralities---such as \textit{Salience}, \textit{Betweenness}, \textit{Degree}, \textit{Strength}, and \textit{PageRank}---produce an immediate impact, swaying the opinions of many agents. 
However, they also create a sharp divide with the rest of the network, which becomes unable to interact with that cluster. 
Consequently, the global attraction process discussed earlier becomes less effective. 
\textit{Random}, \textit{s-coreness}, and \textit{k-coreness} do not exhibit this behavior because their lower inherent influence is balanced by the sheer volume of stubborn agents in the network. 
Among the centrality measures that require network knowledge, \textit{Degree} and \textit{Strength} offer a favorable trade-off between performance and computational cost, as they are static and easily obtainable.
\textit{Random} selection instead, especially in the dynamic setting, delivers comparable results while incurring the lowest implementation cost, as it does not require full network knowledge.

\section{Conclusions}
\label{sec:conclusion}

In this work, we analyzed the effectiveness of attacks that influence the average opinion of a social network. 
We used various centrality measures to select the positions of stubborn agents---i.e., nodes with fixed opinions that attempt to influence their neighbors. 
The Lancichinetti–Fortunato–Radicchi (LFR) network model was chosen for its ability to capture degree heterogeneity and community structure, which are essential characteristics of real social networks. 
For opinion propagation, we adopted the deterministic Hegselmann–Krause continuous model, which allows us to evaluate the impact of each attack and the effectiveness of each centrality measure without the unpredictability of stochastic effects.

Our analysis shows that, under a \textit{Static Strategy} (stubborn agents holding an extreme opinion fixed over time), certain centralities outperform others in identifying optimal target nodes. 
However, due to the bounded confidence mechanism, agents cannot be influenced by neighbors whose opinions differ too greatly. 
Consequently, the fraction of the network that can be ``pulled'' toward the extreme opinion remains limited and tends to saturate as the fraction of stubborn agents increases.

A \textit{Dynamic Strategy} proves far more effective. 
Here, the opinion of stubborn nodes is modulated over time, gradually moving from a moderate value to an extreme one. 
In this scenario, even a very small number of stubborn agents can pull nearly the entire population toward the extreme opinion. 
Moreover, this occurs using virtually any centrality metric---even random placement yields remarkably good performance. 
This is an important observation, given that network structure is often largely unknown.

Of course, this work has limitations, including the use of a single network model and a single opinion dynamics mechanism, although both are widely recognized as representative of realistic situations.
Moreover, our study exclusively considered stubborn agents aiming to push public opinion toward an extreme value. 
In real networks, external agents may have diverse objectives---some may seek moderate shifts, others may attempt to stifle debate or amplify certain interactions. 
Our findings therefore do not directly generalize to all types of opinion conditioning.
Nevertheless, the study lends itself to several future developments. 
For example, the dynamic scenario could be extended to propose alternative strategies and evaluate the optimal trade‑off between effectiveness, speed, and implementation simplicity. 
One could also consider making the conditioning effort dependent on the state of the social network. 
Additionally, the inherent community structure of LFR networks makes them suitable for studying targeted attacks on individual communities, i.e., groups of individuals who presumably share the same opinion.

In conclusion, this work contributes to the research line aimed at understanding opinion propagation in social networks and the mechanisms that can condition it. 
The ultimate goal is to develop effective countermeasures that ensure fairness and impartiality in public debate.


\bibliographystyle{IEEEtran}
\bibliography{bibliography}
%

%





\end{document}


%


\newcommand{\MYtitle}{SUPPLEMENTARY MATERIAL \\Static and Dynamic Strategies for Influencing Opinions in Social Networks}
\title{\MYtitle}
%
%
%
%

\author{Paolo~Tarantino,~Fabio~Mazza,~Carlo~Piccardi,~and Francesco~Pierri
\IEEEcompsocitemizethanks{\IEEEcompsocthanksitem The authors are with the Department
of Electronics, Information, and Bioengineering, Politecnico di Milano, Milan, Italy.\protect\\
Corresponding author: francesco.pierri@polimi.it
}
}

%
%

%

\maketitle

\section{Results for Other Network Sizes ($N=2000$)}

We repeated the experiments described in the main text on LFR networks of size $N=2000$. The results, presented in this Supplementary Material, are qualitatively identical to those obtained with $N=1000$. For brevity, we only include the key figures and summarize the main observations.

\subsection{Static Strategy}

\begin{figure}[!t]
    \centering
    \includegraphics[width=1.\linewidth]{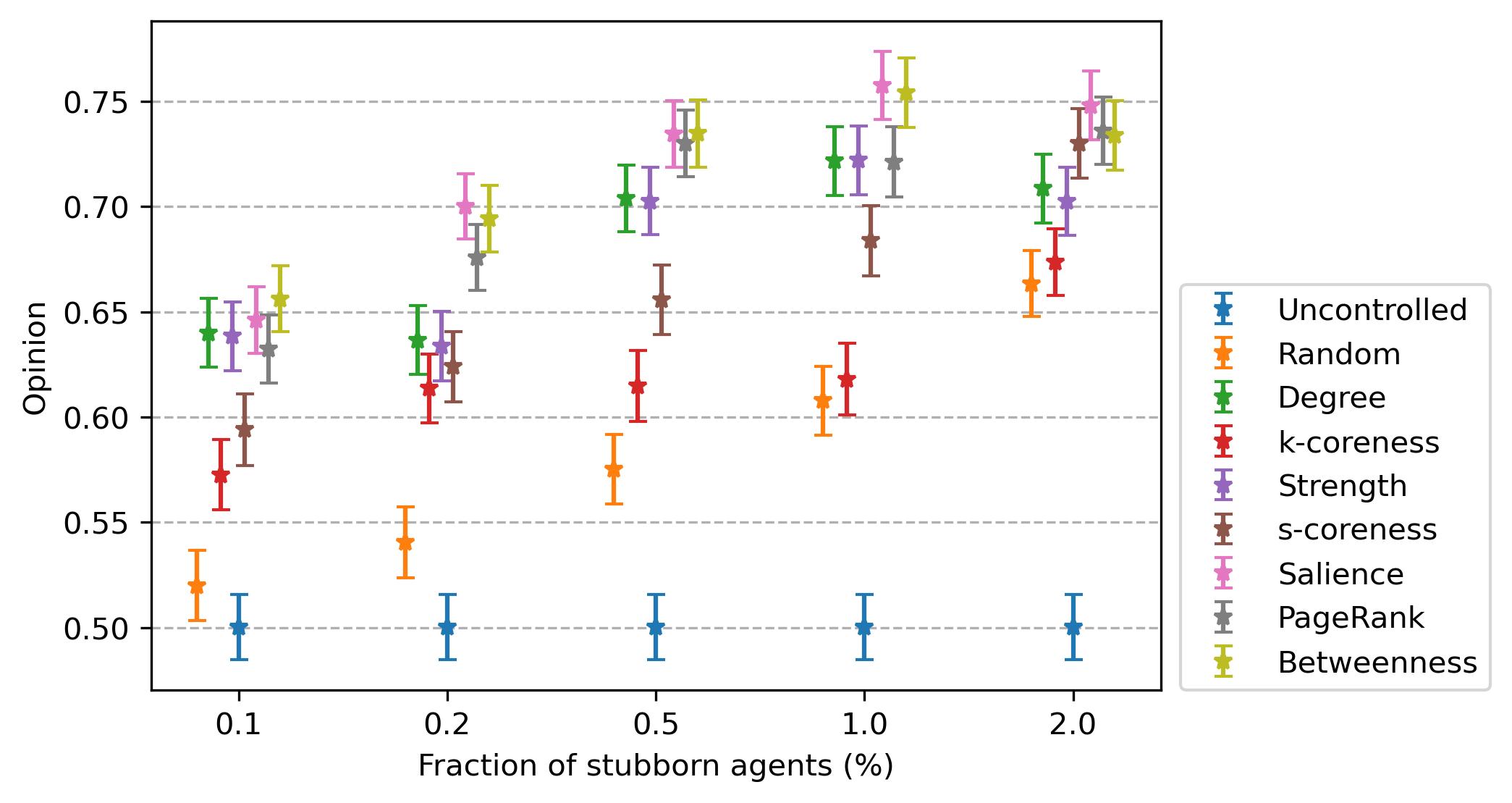}
    \caption{Final average opinion as a function of the fraction $f_S$ of stubborn agents (with fixed opinion $x_S = 1$) and of the centrality used to select them in the \textit{Static Strategy} ($N=2000$).}
    \label{fig:avgbyfrac_opi_1_N2000}
\end{figure}

Figure~\ref{fig:avgbyfrac_opi_1_N2000} shows the final average opinion as a function of the fraction $f_S$ of stubborn agents. As in the $N=1000$ case, centralities such as \textit{Salience}, \textit{Betweenness}, and \textit{PageRank} outperform others, with performance saturating around $f_S = 1\%$.

\begin{figure}[!t]
    \centering
    \includegraphics[width=1.\linewidth]{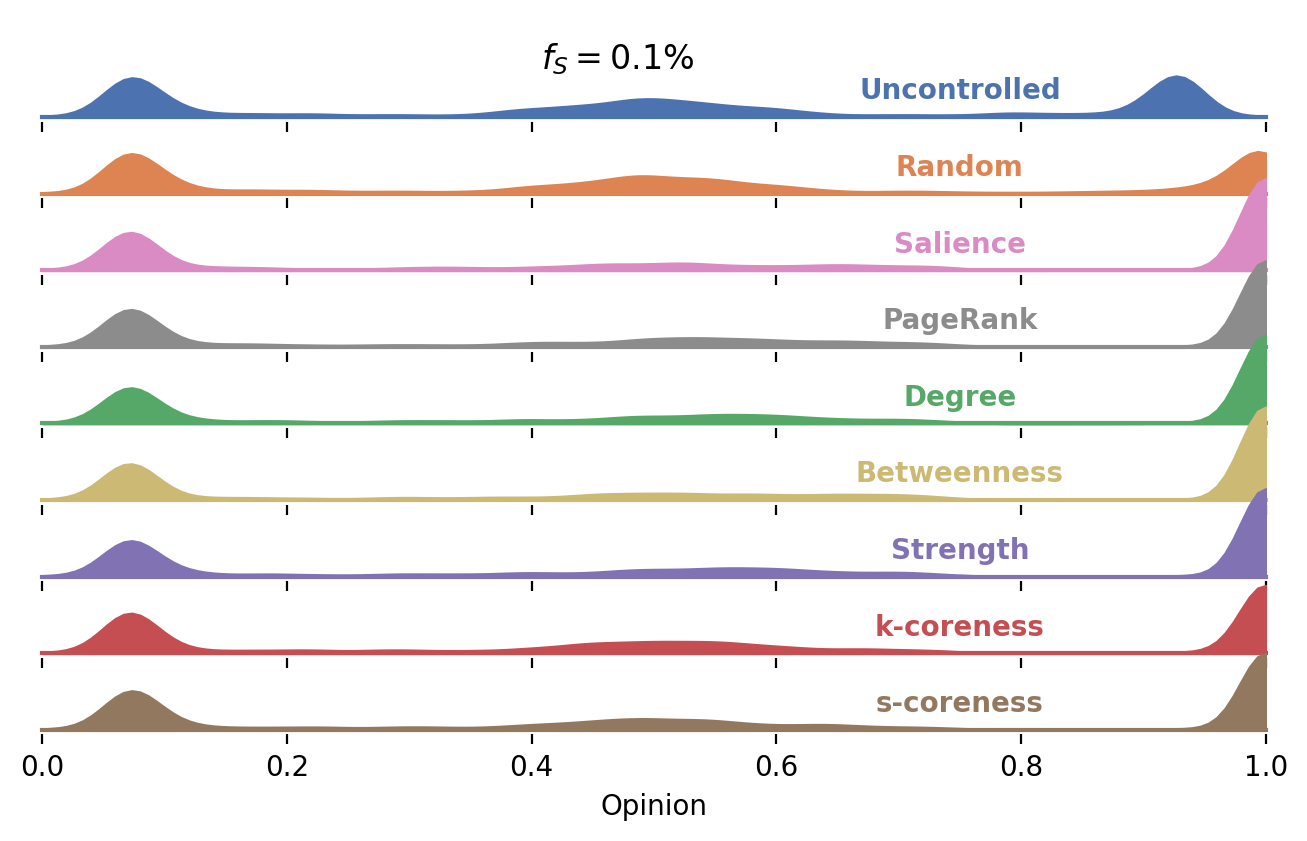}
    \includegraphics[width=1.\linewidth]{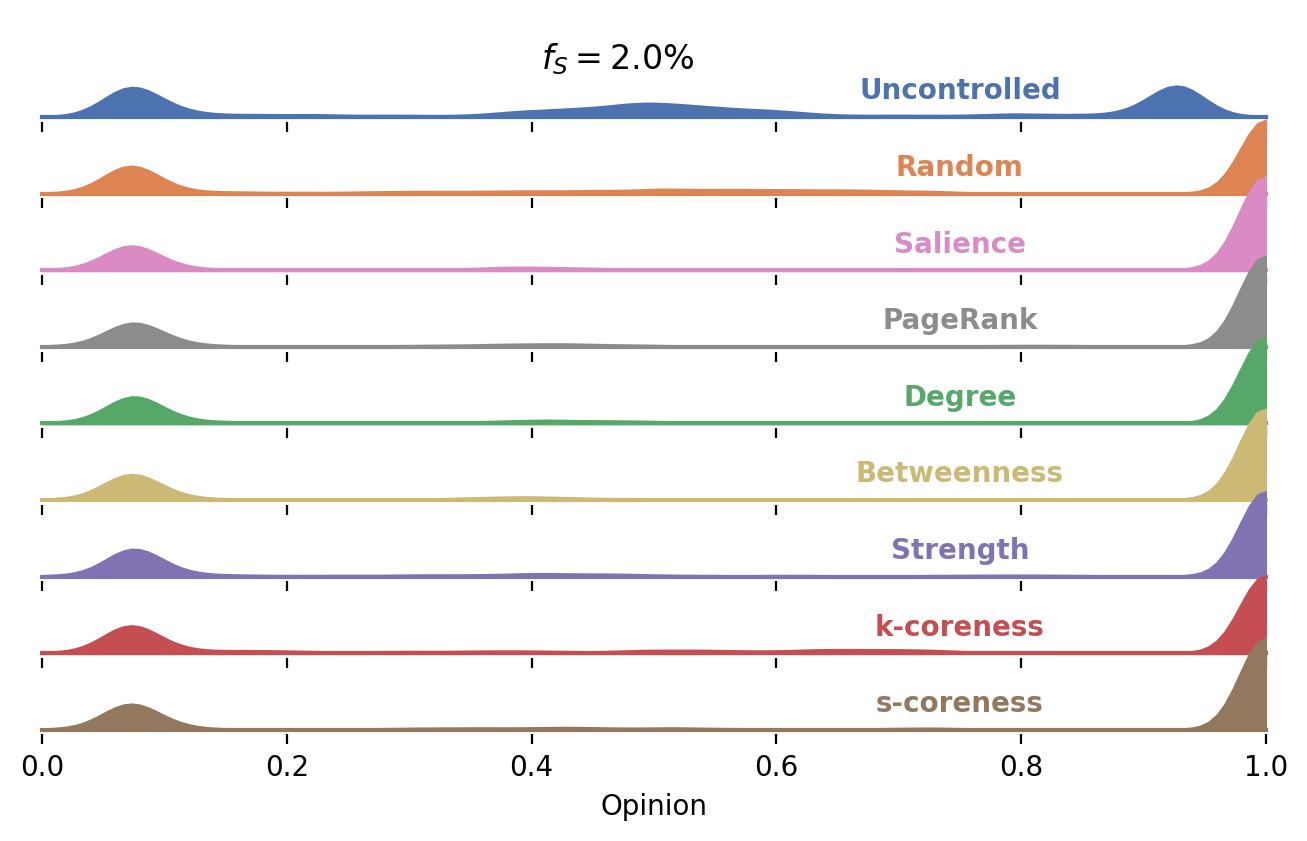}
    \caption{Final opinion distributions for two values of the fraction $f_S$ of stubborn agents (top: $0.1\%$, bottom: $2.0\%$), with fixed opinion $x_S = 1$, obtained for all centrality measures in the \textit{Static Strategy} ($N=2000$).}
    \label{fig:ridge_opi_1_N2000}
\end{figure}

\begin{figure}[p!]
    \centering
    \includegraphics[width=1.\linewidth]{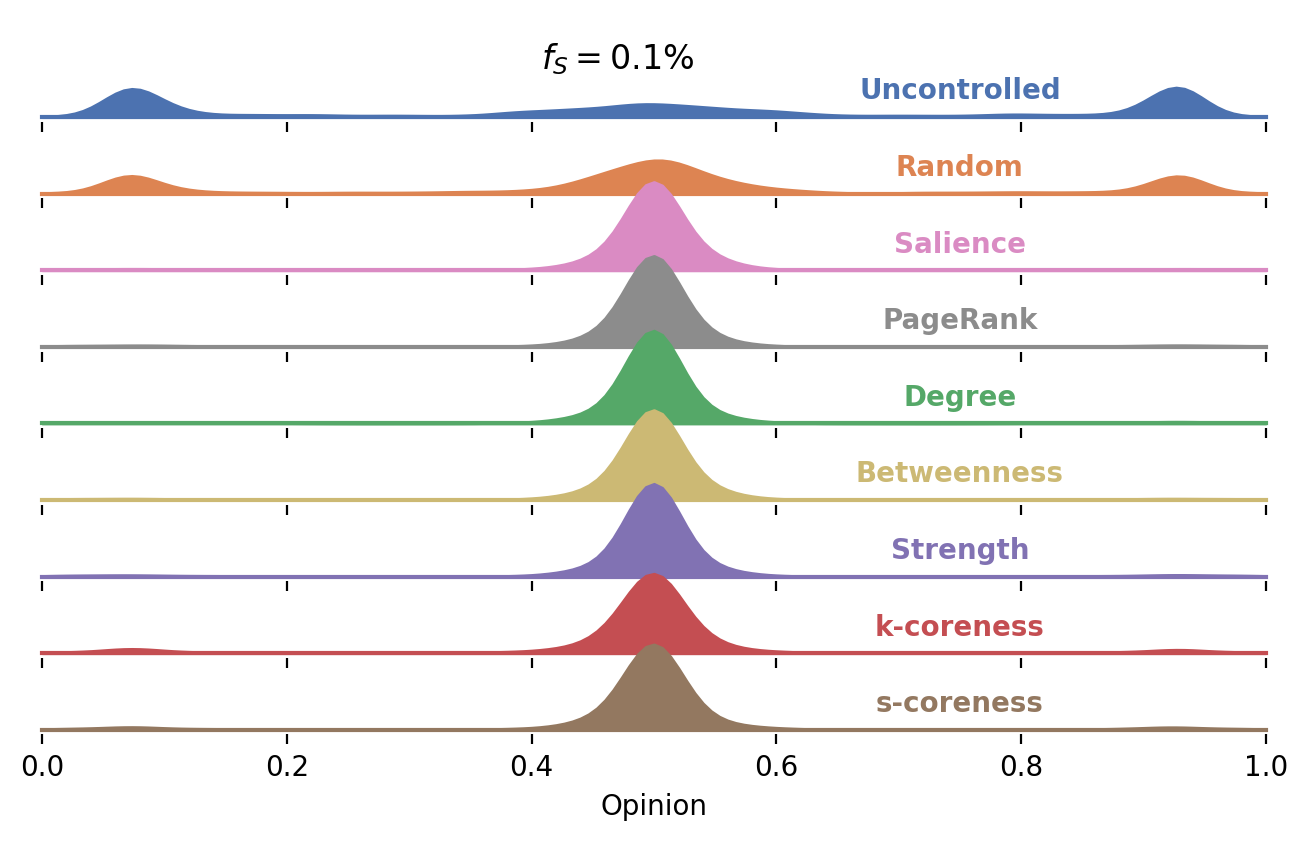}
    \caption{Final opinion distribution for $f_S = 0.1\%$ of stubborn agents with fixed moderate opinion $x_S = 0.5$, obtained for all centrality measures in the \textit{Static Strategy} ($N=2000$).}
    \label{fig:ridge_opi_05_N2000}
\end{figure}

Figure~\ref{fig:ridge_opi_1_N2000} presents the final opinion distributions for $f_S = 0.1\%$ and $f_S = 2.0\%$. The static strategy captures only those agents whose opinions are already close to the target $x_S = 1$, leaving the rest of the distribution largely unchanged. For comparison, Figure~\ref{fig:ridge_opi_05_N2000} shows that setting $x_S = 0.5$ (a moderate opinion) attracts most of the population, as the bounded confidence mechanism does not immediately isolate nodes.

\begin{figure}[!t]
    \centering
    \includegraphics[width=1.\linewidth]{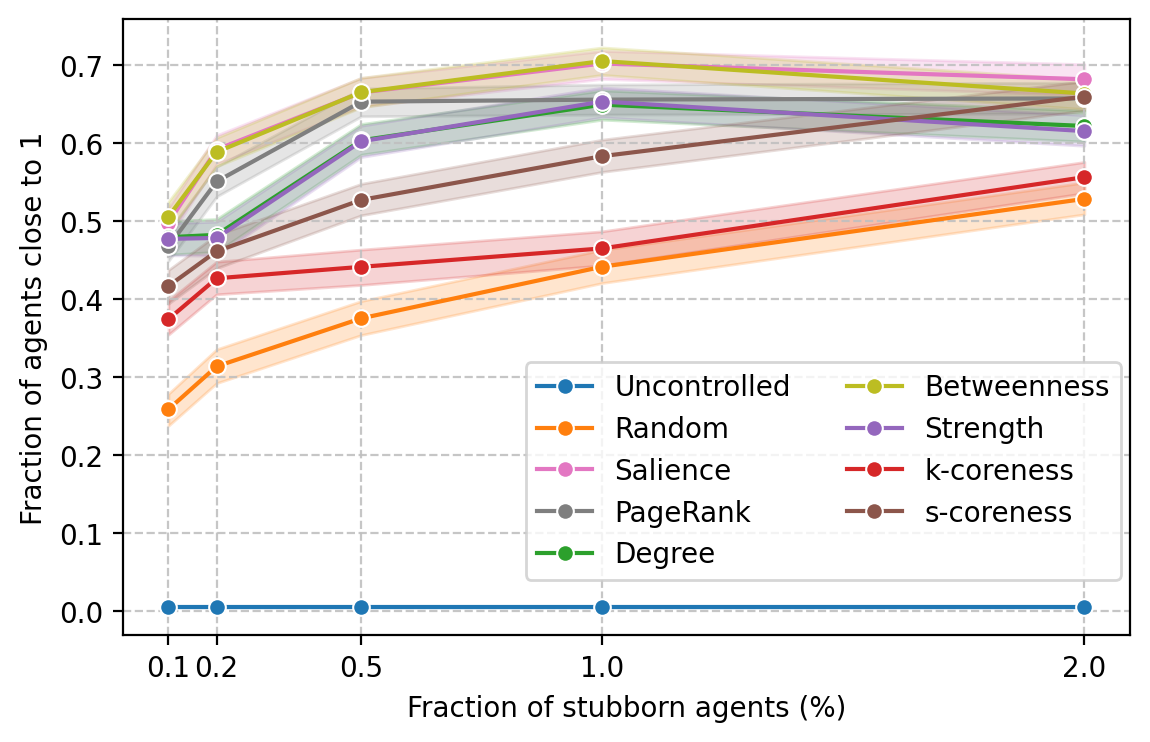}
    \caption{Fraction of the population with final opinion at distance less than $0.05$ from $x_S = 1$, as a function of the fraction $f_S$ of stubborn agents and of the centrality used to select them in the \textit{Static Strategy} ($N=2000$).}
    \label{fig:frac_near_opi_1_N2000}
\end{figure}

\begin{figure}[!t]
    \centering
    \includegraphics[width=1.\linewidth]{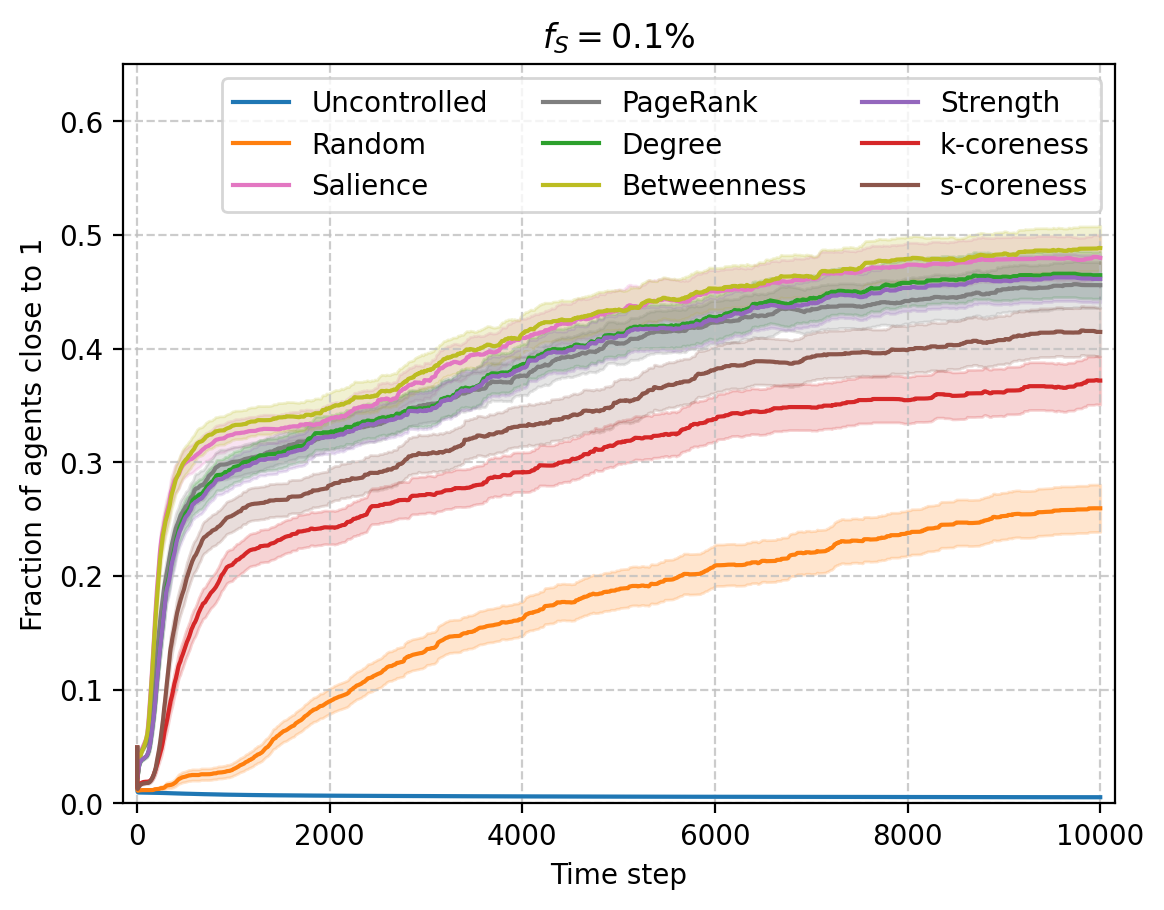}
    \includegraphics[width=1.\linewidth]{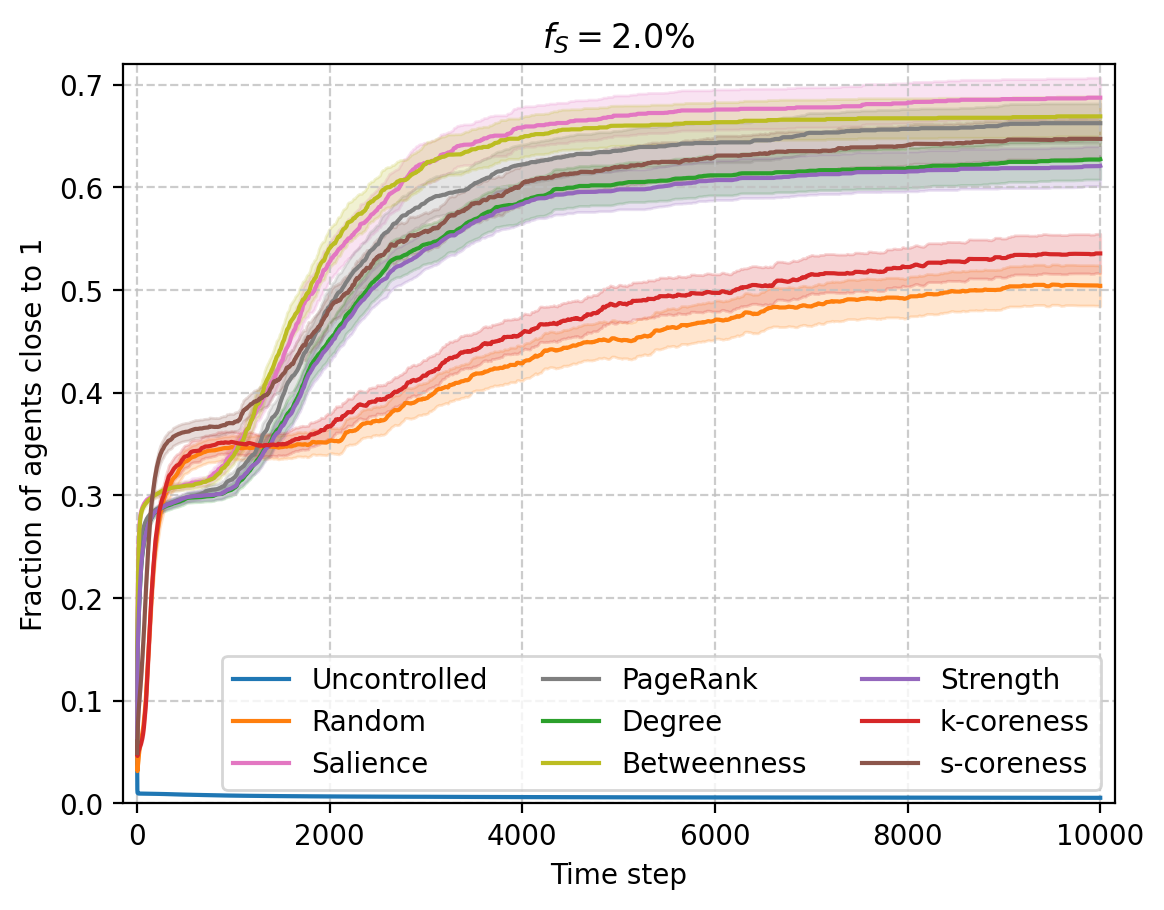}
    \caption{Time evolution of the fraction of the population with opinion at distance less than $0.05$ from $x_S = 1$, for two values of the fraction $f_S$ of stubborn agents (top: $0.1\%$, bottom: $2.0\%$), with fixed opinion $x_S = 1$, obtained for all centrality measures in the \textit{Static Strategy} ($N=2000$).}
    \label{fig:frac_near_hist_opi_1_N2000}
\end{figure}

The fraction of the population with final opinion within $0.05$ of $x_S = 1$ is shown in Figure~\ref{fig:frac_near_opi_1_N2000}. The saturation effect is again evident. Finally, Figure~\ref{fig:frac_near_hist_opi_1_N2000} depicts the time evolution of this fraction for the two extreme values of $f_S$ ($0.1\%$ and $2.0\%$). The initial steep increase, the centrality dependence at low $f_S$, and the subsequent plateau are all consistent with the $N=1000$ results.

\subsection{Dynamic Strategy}

\begin{figure}[!t]
    \centering
    \includegraphics[width=1.\linewidth]{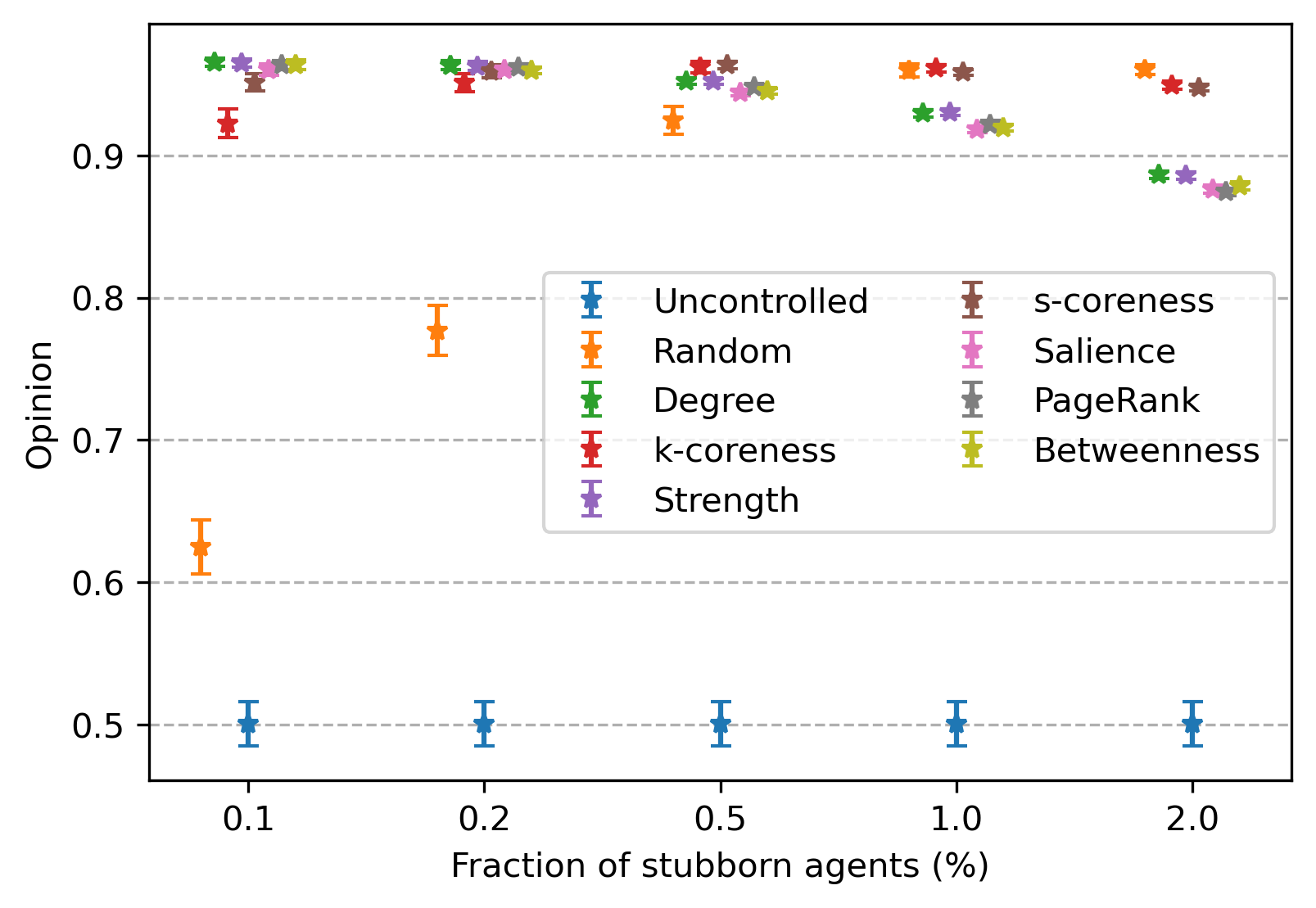}
    \caption{Final average opinion as a function of the fraction $f_S$ of stubborn agents (with final opinion $x_S = 1$) and of the centrality used to select them in the \textit{Dynamic Strategy} ($N=2000$).}
    \label{fig:avgbyfrac_opi_1_N2000_dyn}
\end{figure}

\begin{figure}[!t]
    \centering
    \includegraphics[width=1.\linewidth]{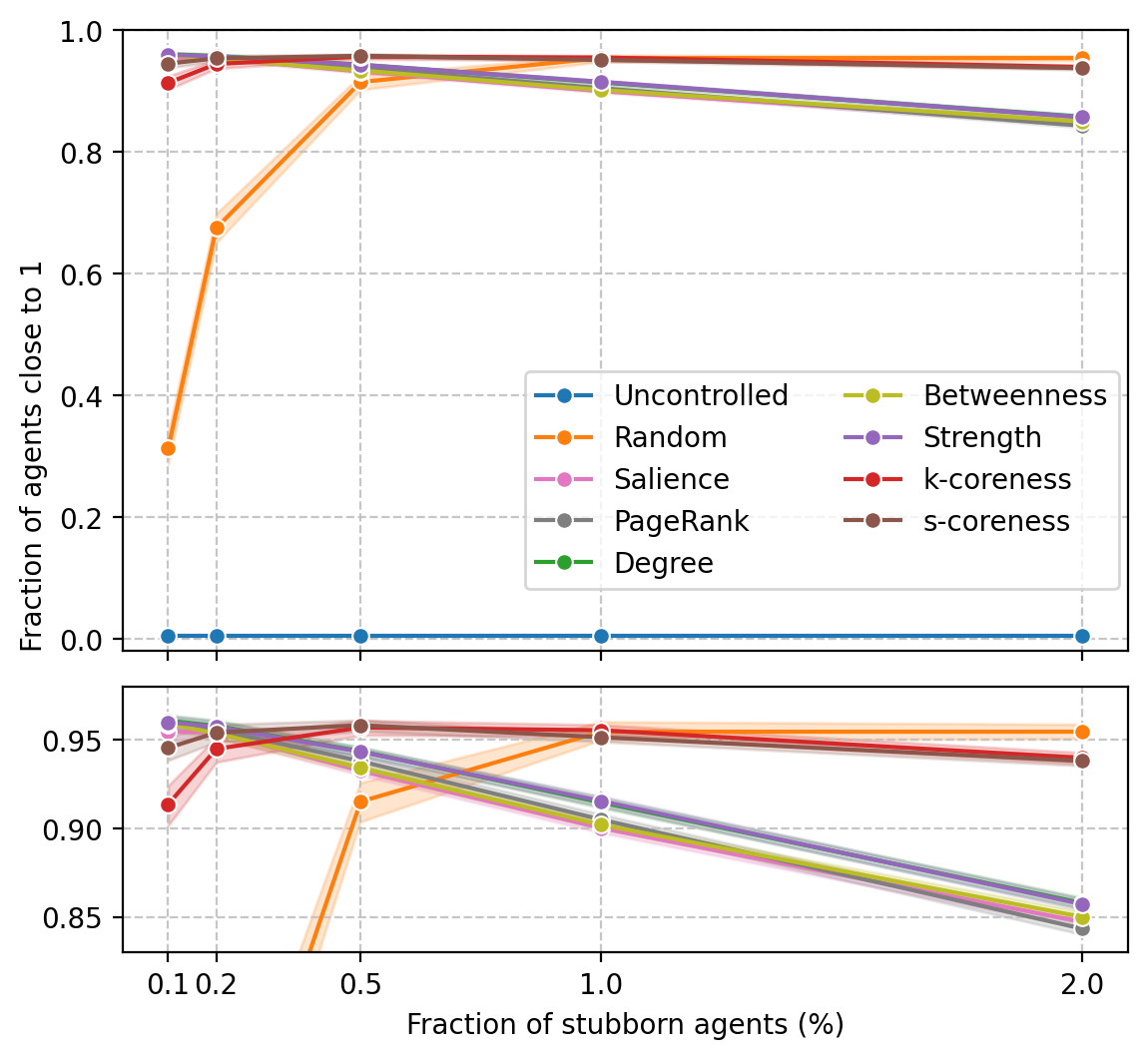}
    \caption{Fraction of the population with final opinion at distance less than $0.05$ from $x_S = 1$, as a function of the fraction $f_S$ of stubborn agents and of the centrality used to select them in the \textit{Dynamic Strategy} ($N=2000$). The bottom panel shows a magnified view of the saturation region from the top panel.}
    \label{fig:dynamic_frac_near_opi_1_N2000}
\end{figure}

For the dynamic strategy, Figure~\ref{fig:avgbyfrac_opi_1_N2000_dyn} reports the final average opinion. The striking effectiveness at low $f_S$ (e.g., $0.1\%$) and the counterintuitive performance reversal at higher fractions are reproduced. Figure~\ref{fig:dynamic_frac_near_opi_1_N2000} shows the fraction of the population close to $x_S=1$ as a function of $f_S$; the bottom panel magnifies the saturation region. As observed for $N=1000$, increasing the fraction of stubborn agents can reduce the attracted population when highly effective centralities are used (e.g., \textit{Betweenness}, \textit{Salience}), whereas \textit{k-coreness}, \textit{s-coreness}, and \textit{Random} selection continue to perform well.

\begin{figure*}
    \centering
    \includegraphics[width=0.85\linewidth]{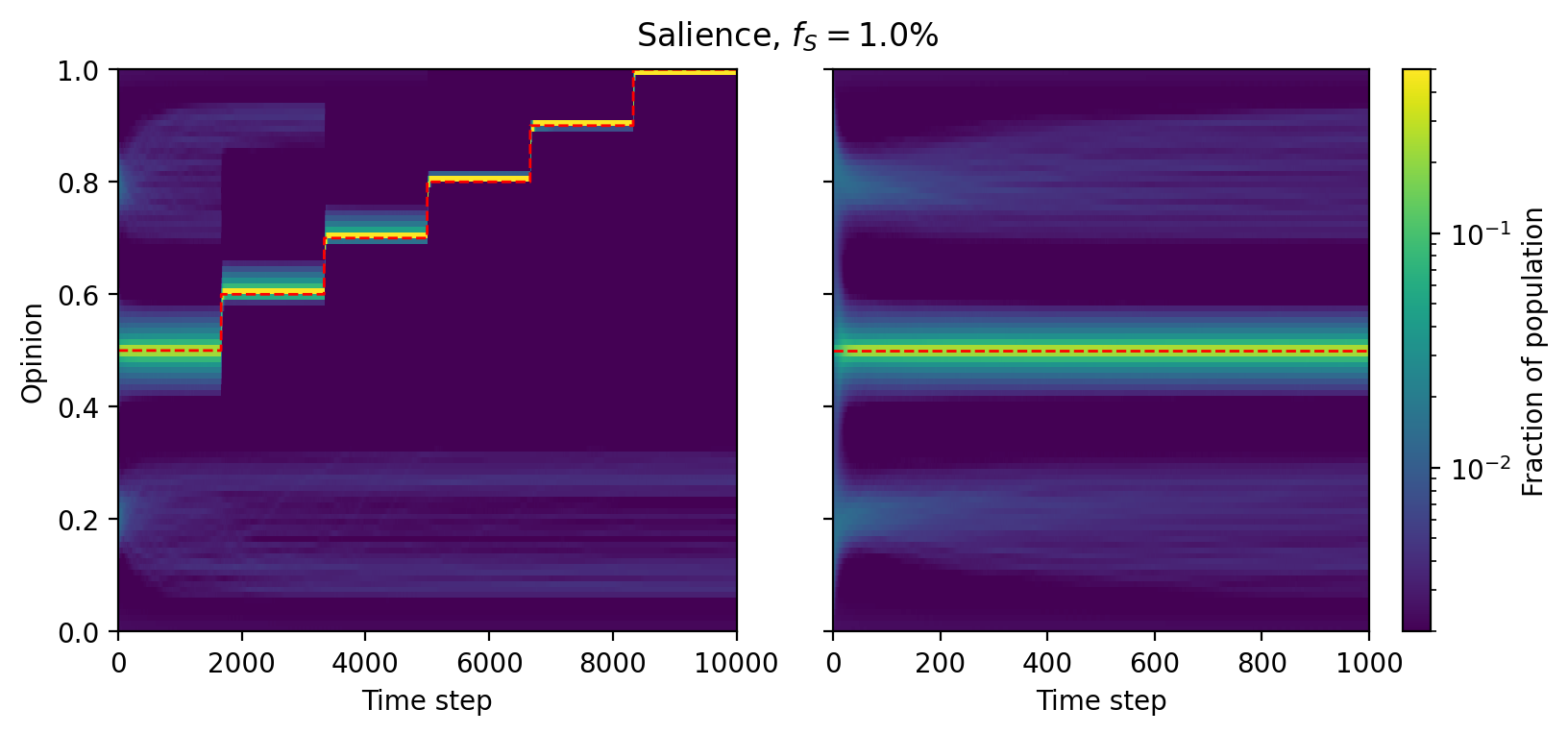}
    \includegraphics[width=0.85\linewidth]{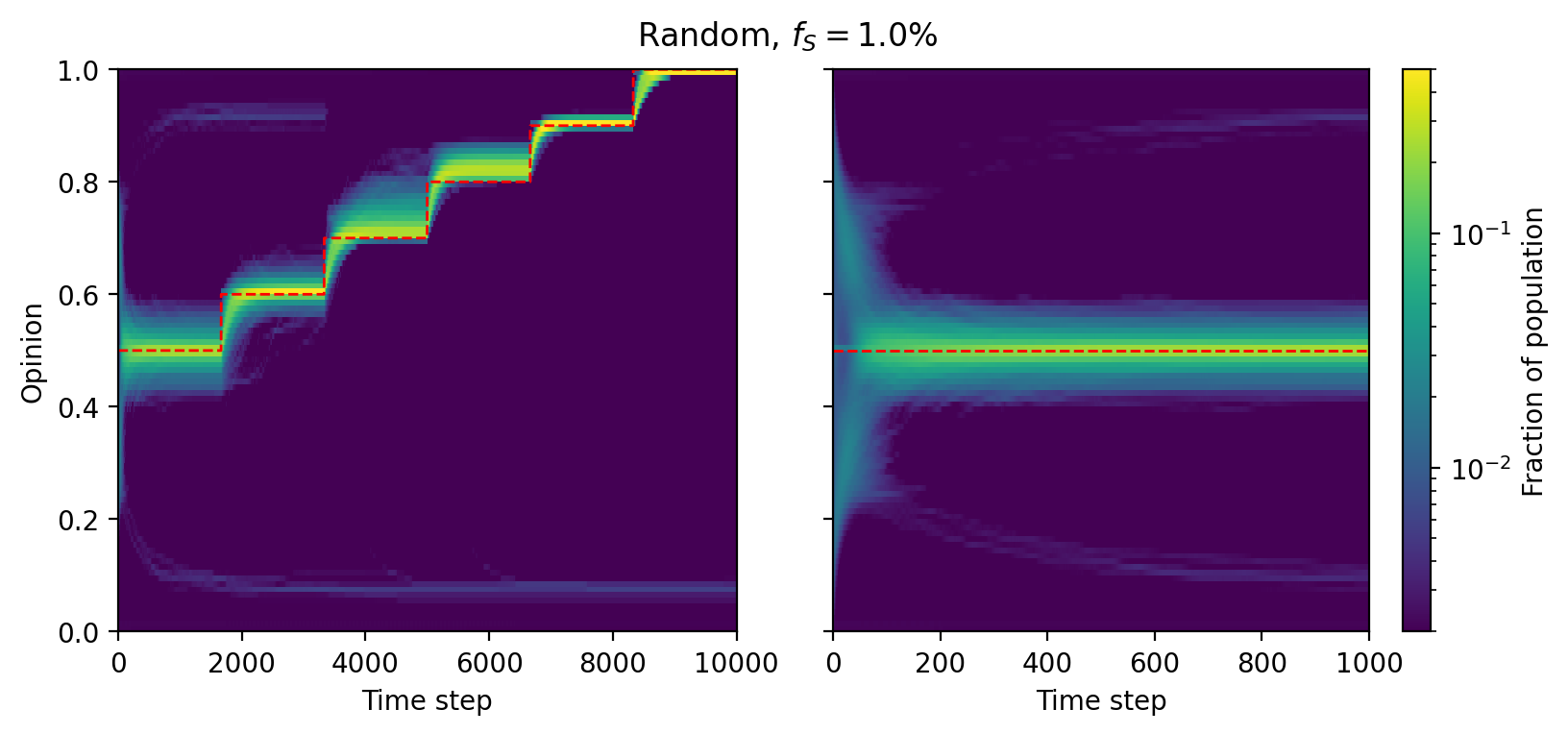}
    \caption{Time evolution of the opinion density, averaged over 50 simulations and 20 network instances, comparing the dynamics obtained with \textit{Salience} (top panels) and \textit{Random} (bottom panels) for a large fraction of stubborn agents ($f_S = 1\%$). The left panels show the entire simulation; the right panels show only the initial stages up to $t = 1000$. The red dashed line indicates the time‑varying opinion of the stubborn agents (\textit{Dynamic Strategy}, $N=2000$).}
    \label{fig:heatmaps_opis_dynamic_2000}
\end{figure*}

Figure~\ref{fig:heatmaps_opis_dynamic_2000} compares the temporal evolution of opinion density for \textit{Salience} (top) versus \textit{Random} (bottom) with $f_S = 1\%$. The rapid lock‑in at $x_S = 0.5$ for Salience, and the gradual cascade toward $x_S = 1$ for Random, mirror the dynamics observed in the main text.

\subsection{Network Statistics}

Tables~\ref{tab:net1000_stats} and~\ref{tab:net2000_stats} report the statistics of the LFR network instances used in our experiments ($N=1000$ and $N=2000$, respectively). For each network, the following quantities are given: $|E|$ (number of edges), $k_{\min}$ and $k_{\max}$ (minimum and maximum degree), $\langle k \rangle$ (average degree), $A$ (assortativity coefficient), $C$ (global clustering coefficient), $\alpha_k$ and $\alpha_w$ (power‑law exponents of the degree and weight distributions).

\clearpage

\begin{table}[p]
    \caption{Statistics of LFR networks, $N=1000$}
    \centering
    \label{tab:net1000_stats}
    \begin{tabular}{lrrrrrrrrr}
        n° & $|E|$ & $k_{\min}$ & $k_{\max}$ & $\langle k \rangle$ & $A$ & $C$ & $\alpha_k$ & $\alpha_w$ \\
        \hline
        1 & 10201 & 6 & 198 & 20.4 & -0.644 & 0.233 & 2.31 & 2.97 \\
        2 & 11140 & 6 & 182 & 22.3 & -0.689 & 0.281 & 2.13 & 2.99 \\
        3 & 9891 & 6 & 199 & 19.8 & -0.587 & 0.221 & 2.24 & 2.79 \\
        4 & 10033 & 6 & 200 & 20.1 & -0.717 & 0.219 & 2.29 & 2.91 \\
        5 & 10512 & 6 & 198 & 21.0 & -0.702 & 0.231 & 2.19 & 3.00 \\
        6 & 10233 & 6 & 199 & 20.5 & -0.689 & 0.226 & 2.24 & 2.86 \\
        7 & 10721 & 6 & 184 & 21.4 & -0.671 & 0.271 & 2.17 & 2.87 \\
        8 & 9613 & 6 & 181 & 19.2 & -0.665 & 0.226 & 2.52 & 3.00 \\
        9 & 10408 & 6 & 198 & 20.8 & -0.656 & 0.235 & 2.24 & 2.71 \\
        10 & 10058 & 6 & 196 & 20.1 & -0.621 & 0.222 & 2.31 & 2.69 \\
        11 & 10050 & 6 & 191 & 20.1 & -0.562 & 0.225 & 2.25 & 2.69 \\
        12 & 10042 & 6 & 177 & 20.1 & -0.533 & 0.256 & 2.21 & 3.00 \\
        13 & 10175 & 6 & 193 & 20.4 & -0.803 & 0.249 & 2.24 & 2.92 \\
        14 & 9237 & 6 & 193 & 18.5 & -0.558 & 0.205 & 2.39 & 2.96 \\
        15 & 10324 & 6 & 199 & 20.6 & -0.587 & 0.225 & 2.36 & 2.90 \\
        16 & 9324 & 6 & 165 & 18.6 & -0.653 & 0.233 & 2.31 & 3.00 \\
        17 & 10840 & 6 & 195 & 21.7 & -0.674 & 0.238 & 2.16 & 2.79 \\
        18 & 10553 & 6 & 198 & 21.1 & -0.662 & 0.251 & 2.20 & 2.26 \\
        19 & 10012 & 6 & 179 & 20.0 & -0.718 & 0.240 & 2.23 & 2.98 \\
        20 & 10066 & 6 & 199 & 20.1 & -0.587 & 0.246 & 2.24 & 2.94 \\
    \end{tabular}
\end{table}

\begin{table}[!t]
    \caption{Statistics of LFR networks, $N=2000$}
    \centering
    \label{tab:net2000_stats}
    \begin{tabular}{lrrrrrrrrr}
        n° & $|E|$ & $k_{\min}$ & $k_{\max}$ & $\langle k \rangle$ & $A$ & $C$ & $\alpha_k$ & $\alpha_w$ \\
        \hline
        1 & 21170 & 6 & 198 & 21.2 & -0.670 & 0.238 & 2.23 & 2.51 \\
        2 & 21612 & 6 & 199 & 21.6 & -0.699 & 0.252 & 2.17 & 2.66 \\
        3 & 20048 & 6 & 199 & 20.0 & -0.611 & 0.233 & 2.24 & 3.00 \\
        4 & 20318 & 6 & 199 & 20.3 & -0.704 & 0.228 & 2.26 & 3.00 \\
        5 & 19981 & 6 & 198 & 20.0 & -0.646 & 0.226 & 2.23 & 3.00 \\
        6 & 20254 & 6 & 199 & 20.3 & -0.698 & 0.240 & 2.22 & 3.00 \\
        7 & 21287 & 6 & 199 & 21.3 & -0.719 & 0.254 & 2.18 & 2.87 \\
        8 & 19698 & 6 & 181 & 19.7 & -0.604 & 0.238 & 2.46 & 3.00 \\
        9 & 21164 & 6 & 198 & 21.2 & -0.711 & 0.237 & 2.18 & 3.00 \\
        10 & 20129 & 6 & 199 & 20.1 & -0.669 & 0.225 & 2.25 & 2.91 \\
        11 & 19848 & 6 & 196 & 19.8 & -0.619 & 0.235 & 2.26 & 2.84 \\
        12 & 20354 & 6 & 199 & 20.4 & -0.669 & 0.234 & 2.22 & 3.00 \\
        13 & 20115 & 6 & 199 & 20.1 & -0.683 & 0.233 & 2.25 & 3.00 \\
        14 & 19751 & 6 & 193 & 19.8 & -0.653 & 0.237 & 2.25 & 3.00 \\
        15 & 20583 & 6 & 200 & 20.6 & -0.707 & 0.235 & 2.30 & 3.00 \\
        16 & 19537 & 6 & 195 & 19.6 & -0.618 & 0.222 & 2.25 & 3.00 \\
        17 & 20917 & 6 & 195 & 20.9 & -0.625 & 0.250 & 2.19 & 2.94 \\
        18 & 21312 & 6 & 198 & 21.3 & -0.668 & 0.248 & 2.19 & 2.61 \\
        19 & 20356 & 6 & 192 & 20.4 & -0.600 & 0.248 & 2.21 & 2.81 \\
        20 & 19714 & 6 & 199 & 19.7 & -0.648 & 0.233 & 2.25 & 3.00 \\
    \end{tabular}
\end{table}
